\documentclass[particles,article,submit,pdftex,moreauthors]{Definitions/mdpi} 
\firstpage{1} 
\pubvolume{1}
\issuenum{1}
\articlenumber{0}
\pubyear{2026}
\copyrightyear{2026}
\datereceived{ } 
\daterevised{ } 
\dateaccepted{ } 
\datepublished{ } 
\usepackage{microtype}
\usepackage{graphicx}
\usepackage{float}
\usepackage{hyphenat}
\newcommand{\req}[1]{Eq.\,({\ref{#1}})}
\newcommand{\reqs}[2]{Eqs.\,({\ref{#1}},{\ref{#2}})}
\pdfoutput=1 
\renewcommand{\linenumbers}{}

\Title{Weinberg Angle, Neutron Abundance in BBN, and Lifetime}
\Author{Cheng Tao Yang\orcidA{}, and Johann Rafelski$^*$\orcidB{} }
\AuthorNames{Cheng Tao Yang, and Johann Rafelski}

\address[1]{Department of Physics, The University of Arizona, Tucson, Arizona 85721, USA\\$^*$ Correspondence: johannr at arizona.edu}

\abstract{We present state of the art kinetic theory determination of the neutron abundance available for the Big-Bang nucleosynthesis (BBN). Our work is motivated by the study of the neutron lifespan measured in the laboratory and the unknown strength of weak interactions coupling constant $G_\mathrm{F}$ at finite temperature in the primordial Universe. We draw attention to the relevant dependence of $G_\mathrm{F}$ on the symmetry breaking Weinberg angle $s^2_\mathrm{W}$, a free parameter in the standard model of particle physics. We establish how the value of $s^2_\mathrm{W}$ by way of $G_\mathrm{F}$ modification influences neutron abundance available for BBN and neutron lifetime.}
\keyword{neutron Abundance in BBN; Weinberg angle; Fermi constant; neutron Lifetime} 
\begin{document}
\section{Introduction}\label{Intro}
The primary motivation for this work is to quantify the primordial neutron abundance available for Big-Bang nucleosynthesis (BBN) using methods of kinetic theory and to study the impact of small weak interaction $G_\mathrm{F}$ variation from present day value. There is continued interest in understanding the cosmological variation of natural constants~\cite{Uzan:2024ded}, with much of recent interest arising in the context of the Hubble tension~\cite{Milakovic:2026hwa}. We studied this context before, exploring the dependence of neutrino freeze-out on the variation of natural constants~\cite{Birrell:2014uka} in the primordial Universe. Our present work extends these considerations to the temperature domain at the edge of BBN nucleosynthesis physics. 

Our motivation to study the neutron primordial abundance derives from a significant dependence of $G_\mathrm{F}$ on Weinberg angle parameter used in the convenient form $s^2_\mathrm{W}\equiv \sin^2\theta_\mathrm{W},$ $ c^2_\mathrm{W}\equiv \cos^2\theta_\mathrm{W}$. $s^2_\mathrm{W}$ is not as rigidly anchored as is the tree level value ~\cite{Martin:2025cas} $G_\mathrm{F}^\mathrm{tree}={1}/{\sqrt{2}v^2} $ only dependent on the Higgs vacuum expectation value $\langle h\rangle\equiv v$, where $v=246.2$\,GeV and $1/\sqrt{2}v^2=1.1665\times10^{-5}\,\mathrm{GeV}^{-2}$. This is the dependence we considered in the prior study of natural constants~\cite{Birrell:2014uka}. 

Allowing for higher radiative corrections, $G_\mathrm{F}$ and thus the relevant weak interaction (WI) reaction rates become dependent on $s^2_\mathrm{W}$, a free parameter of the Standard Model (SM) of particle physics. We consider in Section~\ref{EWparameters} how the relevant to nucleosynthesis primordial Universe variation in $G_\mathrm{F}$ could originate in a shift of $s^2_\mathrm{W}$. We apply the same method to study the dependence of the neutron lifetime on this symmetry-breaking parameter of electroweak interactions.
 
The well studied radiative vacuum polarization type correction~\cite{Veltman:1977kh,ParticleDataGroup:2024cfk} to the charged current mediating $W$-Boson. The leading order result is~\cite{Martin:2025cas} 
 \begin{align}\label{GFform1}
G_\mathrm{F} = \frac{1}{\sqrt{2}v^2} \left(1+\Delta\alpha\!-\frac{c^2_\mathrm{W}}{s^2_\mathrm{W}}\Delta\rho+\Delta r_\text{rem}\right)\,,
\end{align}
where $\Delta\alpha$ is the fine-structure constant radiative correction, $\Delta\rho$ is the charged current gauge $W$-boson vacuum polarization correction, and $\Delta r$ includes sub-leading corrections. 

In Section~\ref{SiW} we look closer at the Weinberg angle, discuss its possible and relevant for this work variation, possibly originating in ambient properties such as temperature and external fields. In Section~\ref{Fermi} we discuss further the Fermi constant and its electro-weak radiative corrections, highlighting its dependence on the Weinberg angle. We show, somewhat unexpected, how a small variation in $s^2_\mathrm{W}$ through its effect on $G_\mathrm{F}$ modifies in a noticeable manner WI reaction rates.

We review the theory and experiment on neutron lifetime in Section~\ref{nlifetime}. In Section~\ref{DecayMedium} we study how the primordial plasma medium in the Universe modifies the neutron lifetime by blocking the decay phase space. Combining this with the thermal model and neutron decay during the 200 seconds to reach the onset of BBN at $T\approx0.080\,\mathrm{MeV}$, one obtains a first view of expected neutron abundance in the Universe~\cite{Bernstein:1988ad}.

The study of Fermi blocking introduced the general method of approach, allowing the study of neutron and proton reactions on the lepton Universe background, see Section~\ref{WInpRates}. The freeze-out of neutron abundance turns out to require a study of the kinetic population equation in order to understand precisely the neutron abundance that we address in Section~\ref{Neutron}. This more refined understanding of the role of electro-weak processes in a kinetic model allows us to quantify the dependence of the neutron abundance on the actual primordial value of the Weinberg angle. 

To obtain precise neutron abundance we need to have a precise understanding of the ratio of weak interaction strength to the Hubble expansion rate $H$, which we discuss in Appendix~A based on our recent work~\cite{Rafelski:2024fej}. We show the precise dependence of $H$ on the baryon and entropy content, usually stated in form of baryon to photon ratio $\eta_b$, and the number of neutrino species $N_\nu$, which is near but above the value three, since by definition it comprises an inflow of entropy due to pre-freeze-out neutrino reheating by electron-positron annihilation.
We conclude this work with discussion of the implications of our results in Section~\ref{Discussion}.

\section{Electroweak Theory Parameters}\label{EWparameters}
\subsection{Weinberg angle} \label{SiW}
In some beyond standard model (BSM) theoretical models, the magnitude of the electro-weak symmetry breaking is predicted and thus the value of $s^2_\mathrm{W}$ is fixed. This is not the case in SM, where $s^2_\mathrm{W}$ is a free parameter. The value may be associated with the lowest vacuum energy configuration. However, such a theoretical determination awaits. 

The value of $s^2_\mathrm{W}$ controls the splitting in mass of the gauge Bosons $W,Z$ and yet it appears in the WI dominating charged current interactions mediated by $W$ only in beyond the tree level radiative corrections, see~\req{GFform1}. Given this situation we expect that any future dynamic theory determining the value of $s^2_\mathrm{W}$ by seeking the minimum of an effective action includes similar in magnitude effects of both charged and neutral currents contributions to the vacuum energy. At the time of writing, there is nothing in literature we could find that would offer evaluation of radiative corrections as the origin of the value of the Weinberg angle and no reference could be found addressing the temperature dependence. All work which followed the foundational in this context work of Coleman and Weinberg~\cite{Coleman:1973jx}, and the companion papers establishing finite-temperature symmetry behavior~\cite{Dolan:1974qd,Weinberg:1974hy}, appear to address unrelated questions.

Since these questions are not yet addressed in the literature we advance intuitive arguments: Given the radiative dependence of $G_\mathrm{F}$ on $s^2_\mathrm{W}$, the minimum in effective potential as a function of $s^2_\mathrm{W}$ obtained at radiative level could be governed by much lower energy scale compared to the usual in the SM. Seeing the required radiative factors $\alpha^{2n}$, this is so. Therefore the value of the Weinberg angle and thus $G_\mathrm{F}$ would be unusually sensitive to the environmental conditions including temperature, and external (strong) fields. Viewed from another perspective: Symmetry breaking in quantum mechanics is predominantly a function of the environmental conditions. Elementary examples of symmetry breaking in atomic physics by external fields are the Zeeman effect and the Stark effect. 

The experimental value of the Weinberg angle is discussed in Refs.\,\cite{Sirlin:2012mh,ParticleDataGroup:2024cfk}, $s^Z_\mathrm{W}\equiv \sin^2\theta(m_Z)_\mathrm{W}=0.2313$. This value relates to experimental data obtained in experiments near to the indicated scale $m_Z$, the mass of the $Z$-gauge meson mediating neutral current. Yet a somewhat larger value $s^{2,0}_\mathrm{W}=0.238$ for a lower energy scale is suggested by recent global result analysis~\cite{Gwinner:2022ijo}. The more recently reported LHCb~\cite{LHCb:2024ygc} and CMS~\cite{CMS:2024ony} results agree. The scale dependence (running) of $s^2_\mathrm{W}$ at low energies is studied in~\cite{Erler:2004nh2}. However, the recommended NIST-CODATA value~\cite{NIST-CODATA} we adopt in this work addressing low energy phenomena is smaller, $s^2_\mathrm{W}=0.223$. The difference between these two low energy values is nearly 7\%. Such value scatter can be interpreted as a combination of several effects, including differences in renormalization scheme and unresolved experimental systematics. However, it is possible to question whether a small additional dependence of as a yet unidentified physical origin could be present. One such possibility, which we wish to highlight for future study, is that the unconstrained free parameter of the SM, $s^2_\mathrm{W}$, might exhibit a hidden small residual sensitivity to environmental factors such as external electromagnetic fields. In this context we note that a much smaller than 7\% variance in $s^2_\mathrm{W}$ suffices to explain the disagreement in the measured values of neutron lifetime. 

It is possible that the value of the symmetry breaking Weinberg angle can change at percent-level for temperature as low as a fraction of a MeV given the theoretical considerations, without contradicting any experimental result. This is so, since, as discussed, the minimum in vacuum energy as a function of the Weinberg angle is probably flat. Therefore it seems appropriate to explore the sensitivity of neutron abundance at the edge of BBN to modest modification of $s^2_\mathrm{W}$, and also to motivate the future study of the SM effective potential as a function of $s^2_\mathrm{W}$. We emphasize that the arguments offered here are intuitive and therefore only provide motivation; a theoretically complete electroweak determination of the temperature dependence of $s^2_\mathrm{W}$ remains to be done. The present work is a phenomenological sensitivity study within the uncertainty range of $s^2_\mathrm{W}$. We use $s^2_\mathrm{W}$ as an external parameter, vary it within a range bracketing the current experimental scatter, and quantify the resulting impact on the BBN-onset neutron abundance.

More generally, the finite temperature could, in principle, influence the value of the vacuum expectation value $v$ of the Higgs field. This requires temperatures which are above 20 GeV, as we will address in a quantitative manner in the near future. Computation of this effect is possible since action of the SM has a well defined minimum as a function of $v$. As discussed, we do not know the shape of the minimum in vacuum energy as a function of the Weinberg angle and a dynamical theory was not developed.

To conclude: 
\begin{itemize}
\item We believe that the effective electro-weak vacuum energy minimum as a function of $s^2_\mathrm{W}$ is relatively shallow rendering the value susceptible to modification by environmental factors;
\item In the primordial Universe the effective value of the Weinberg angle $s^2_\mathrm{W}$
can differ from its low-energy laboratory value due to the thermal plasma background effects. As a result, the Fermi constant $G_\mathrm{F}$ in this environment could acquire a temperature dependent shift.
\item We consider it possible that value of the Weinberg angle $s^2_\mathrm{W}$ is different in different experimental environments in the presence of strong electromagnetic fields, which could result in method dependent neutron lifetime measurement.
\end{itemize}

\subsection{Fermi constant: Radiative corrections}\label{Fermi}
In this section we explore the sensitivity of the effective Fermi constant to variations in the Weinberg angle. We work within the on-shell renormalization scheme set out by Sirlin and collaborators~\cite{Sirlin:1980nh,Sirlin:2012mh,ParticleDataGroup:2024cfk}, in which the independent input parameters are taken as $\alpha$, $G_\mathrm{F}$, $m_Z$, and the fermion masses, with $s^2_\mathrm{W}$ defined through the on-shell relation $s^2_\mathrm{W} = 1 - m_\mathrm{W}^2/m_Z^2$. In the parametric exploration that follows, we hold the Higgs vacuum expectation value $v$, the fine-structure constant $\alpha$, the top-quark coupling $g_t$, and the top-quark mass $m_t$ fixed, and vary $s^2_\mathrm{W}$ as an external parameter. What follows should therefore be read as a phenomenological tool quantifying how $G_\mathrm{F}$ would respond if $s^2_\mathrm{W}$ were genuinely shifted from its CODATA value; we do not provide a derivation of environment-dependence from first principles. Within the Standard Model, the most precise value of the Fermi constant~\req{GFform1} was obtained from precisely measured muon lifetime~\cite{MuLan:2012sih} $G_\mathrm{F}^\mu
=1.1663787(6)\times10^{-5}\,\mathrm{GeV}^{-2}$. $G_\mathrm{F}$ can be written using measured quantities in the form~\cite{Sirlin:1980nh, Sirlin:2012mh}
\begin{align}\label{GFform2}
 G_\mathrm{F} =\frac{\pi\alpha}{\sqrt{2}m_\mathrm{W}^2s^2_\mathrm{W}}\left(1+\Delta r\right)\,,\quad 
 \mathrm{where\ in\ SM}\quad 
 e^2=g^2 s^2_\mathrm{W},\ m_\mathrm{W}^2=g^2 \frac{v^2}{4}\,.
\end{align}
In the temperature range considered here, $1>T>0.01$\,MeV, we take the Higgs field vacuum expectation value $v=246.2$\,GeV to be a constant. The radiative corrections are summarized in the quantity $\Delta r$. The quantum correction $\Delta r$ has seen extensive theoretical study. The one-loop correction can be written as~\cite{ParticleDataGroup:2024cfk, Erler:2004nh}
\begin{align}
 \Delta r=\Delta\alpha-\frac{c^2_\mathrm{W}}{s^2_\mathrm{W}}\Delta\rho+\Delta r_\text{rem},
\end{align}
where $\Delta\alpha$ is the correction due to the running of fine-structure constant $\alpha$; we have
\begin{align}
 \Delta\alpha=\frac{\hat\alpha(m_Z)-\alpha}{\hat\alpha(m_Z)}\approx0.06654.
\end{align}
The second term $\Delta\rho$ quantifies how neutral-current and charged-current interactions are modified by the loop effect. The leading contribution to $\Delta\rho$ arises from the mass splitting between the top and bottom quarks and depends quadratically on the top-quark mass, we have~\cite{Consoli:1989fg}
\begin{align}
 \Delta\rho=\frac{3\,G_\mathrm{F}m^2_t}{8\sqrt{2}\pi^2}=\frac{3\,G_\mathrm{F}}{8\sqrt{2}\pi^2}\left(\frac{g_tv}{\sqrt{2}}\right)^2,
\end{align}
where we rewrite the top quark mass in terms of vacuum expectation value; we have $m_t=g_t v/\sqrt{2}$ with top quark coupling constant $g_t\approx 1$.
The remainder term $\Delta r_\text{rem}$ contains sub-leading contributions, including the dependence on the Higgs-boson mass. Numerically, its impact is small, contributing at the $\sim 1\%$ level~\cite{Erler:2004nh}. 

In the following calculation, the remainder term is negligible; we thus obtain a self-consistent Dyson-like equation for the Fermi constant:
\begin{align}\label{GFcorr}
 G_\mathrm{F} = \frac{1}{\sqrt{2}v^2}\!\left(1+\Delta\alpha\!-\!\frac{c^2_\mathrm{W}}{s^2_\mathrm{W}}\frac{3G_\mathrm{F}}{8\sqrt{2}\pi^2}\frac{v^2}{2}\right).
\end{align}
We can solve in a self-consistent manner 
\begin{align}
G_\mathrm{F}^\mathrm{sum}=\frac{1}{\sqrt{2}v^2}\left(\frac{1+\Delta\alpha}{1+3c^2_\mathrm{W}/32\pi^2s^2_\mathrm{W}}\right)\,.
\end{align}
However, this is an infinite series and very likely higher order terms are not consistently accounted for. Therefore we will consider the lowest order perturbative modification 
\begin{align}
G_\mathrm{F}^{\mathrm{1st}}\approx\frac{1}{\sqrt{2}v^2}\left({1+\Delta\alpha}-\frac{3c^2_\mathrm{W}}{32\pi^2s^2_\mathrm{W}}\right).
\end{align}
From experimental measurements, the experimental Fermi constant value is defined at the measured value of $s^2_\mathrm{W}$
\begin{align}
G_\mathrm{F}^\mathrm{exp}=\left.G_\mathrm{F}\right|_{s^2_\mathrm{W}=0.223 (\mathrm{measured})}\,.
\end{align}
\begin{figure}
\begin{center}
\hspace*{-0.4cm} \includegraphics[width=5.in]{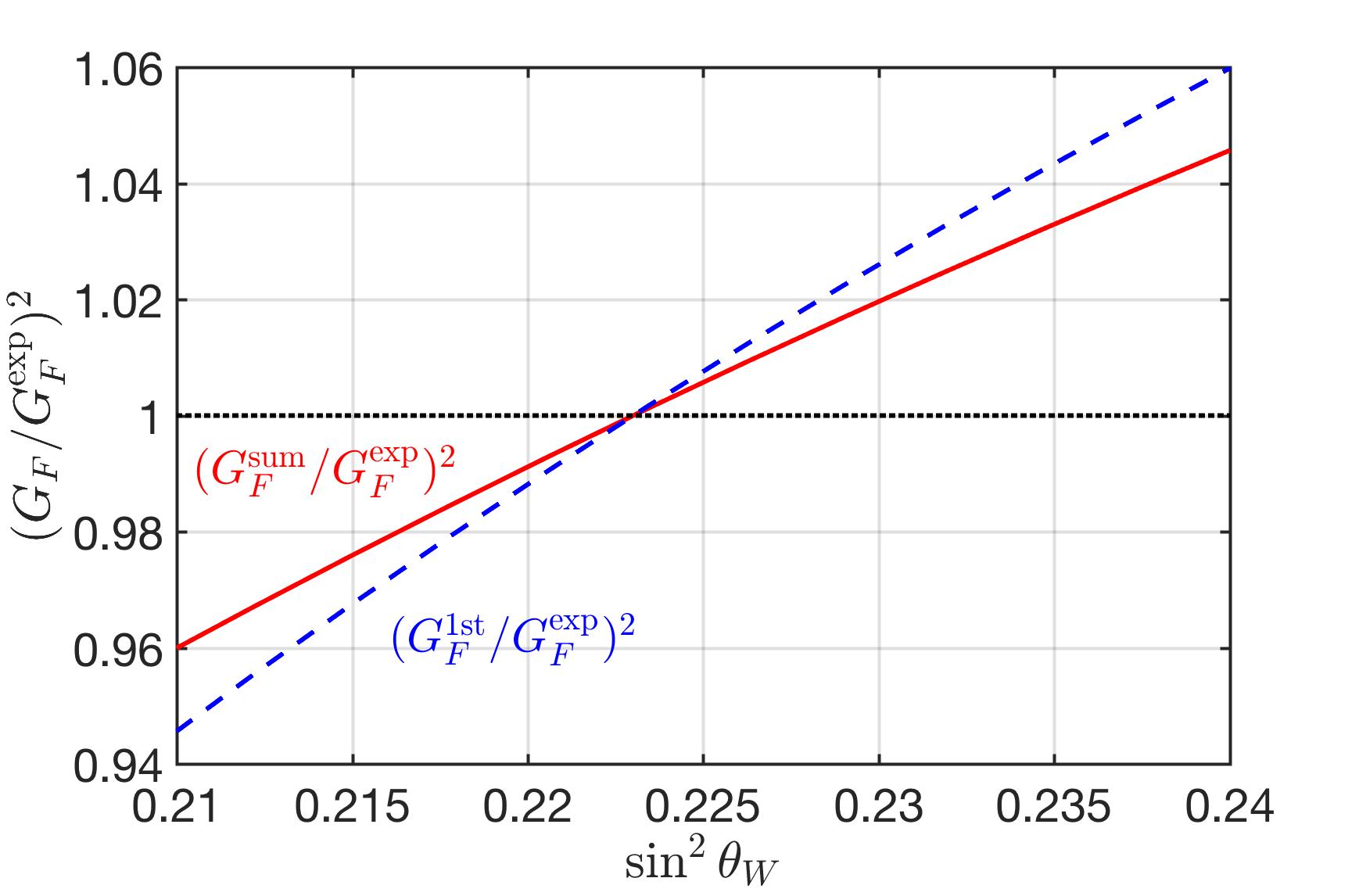}
\caption{The normalized Fermi constant $(G_\mathrm{F}/G_\mathrm{F}^\mathrm{exp})^2$ as a function of the Weinberg angle $s^2_\mathrm{W}$ within the range $0.21<s^2_\mathrm{W}<0.24$. The horizontal dotted line indicates the standard-model value $G_\mathrm{F}=G_\mathrm{F}^\mathrm{exp}$. The two methods of treating the radiative corrections are discussed in text.}
\label{GFWeinberGFig}
\end{center}
\end{figure}
In Fig.~\ref{GFWeinberGFig}, we quantify the dependence of the normalized Fermi constant $(G_\mathrm{F} / G_\mathrm{F}^{\mathrm{exp}})^2$ on the Weinberg angle in the range $0.21 < s^2_\mathrm{W} < 0.24$. The impact of even small variations
in the Weinberg angle on the Fermi constant $G^2_F$ is non-negligible. We note that the presented range of $s^2_\mathrm{W}$ can account for the observed sub-percent discrepancy in neutron lifetime. This demonstrates that precision measurements of neutron decay are sensitive probes of electro-weak parameters and that even slight shifts in the Weinberg angle could have measurable effects on weak interaction processes.

\section{Neutron Decay Rate}
\subsection{Decay in vacuum}\label{nlifetime}

For the neutron decay channel
\begin{align}\label{Ndec}
n\longrightarrow p+e+\overline{\nu}_e\;,
\end{align}
the vacuum lifetime of neutron can be written as:
\begin{align}\label{VacNdecay}
\frac{1}{\tau_n^0}= \frac{1}{2m_n}\int\frac{d^3p_{\bar{\nu}}}{(2\pi)^32E_{\bar{\nu}}}\frac{d^3p_p}{(2\pi)^32E_p}\frac{d^3p_e}{(2\pi)^32E_e} (2\pi)^4\delta^4\left(p_n-p_p-p_e-p_{\bar{\nu}}\right)\langle|\mathcal{M}|^2\rangle,
\end{align}
where the invariant matrix element for the neutron decay for non-relativistic neutron and proton is given by
\begin{align}
\langle|\mathcal{M}|^2\rangle\!\approx\!16\,G^2_FV^2_{ud}\,m_nm_p(1\!+\!3g^2_A)E_{\bar{\nu}}E_e.
\end{align}
 
The neutron lifetime is seen to be proportional to several parameters; the textbook final result is~\cite{Czarnecki:2018okw,Marciano:2014ria,Marciano:2005ec,Czarnecki:2004cw}
\begin{align}\label{eq:taunqual}
&\Gamma^0_{n}=\frac{G^2_F V^2_{ud}Q^5}{2\pi^3}(1+3g^2_A)\mathcal{F}\,f =D\,\mathcal{F}\,f\,,
\end{align}
where the phase space factor $f $ is 
\begin{align}
f \equiv \int^1_{m_e/Q} d\xi\,{\xi(1-\xi)^2} \sqrt{\xi^2 - \left(\frac{m_e}{Q}\right)^{\!\!2}} = 0.01575\,. 
\end{align}
This factor $f$ varies from reaction to reaction and in medium; hence we separated out the reaction strength $D$ appearing here and below in all (thermal) reaction rates allowing to write
\begin{equation}\label{DimFac}
\Gamma^\prime=\Gamma/D\,,\qquad 
D=
\frac{G^2_F V^2_{ud}Q^5 }{2\pi^3}\,(1+3g^2_A)\,.
\end{equation}
In our numerical results 
the strength $D$ is fixed by $D\,\mathcal{F}f=1/\tau_n^0$ with $\tau_n^0=878.4$\,s, which absorbs the radiative correction. $Q$ is the neutron-proton mass difference
\begin{align}\label{eq:Qmdif}
Q=m_n-m_p=1.2933324(5) \,\mathrm{MeV}\,.
\end{align}
When more than 5 digits precision are needed, the more accurate form reads
\begin{align}\label{eq:QmdifEx}
Q=(m_n-m_p)\left[1- \frac{m_n-m_p}{2m_n}\left(1-\frac{m_e^2}{(m_n-m_p)^2}\right)\right]\,.
\end{align}

The nucleon structure enters~\req{eq:taunqual} in terms of $g_A=1.2755$, the ratio of transition axial current to transition vector current. The electro-weak structure is encoded in the weak Fermi coupling $G_\mathrm{F}$ and $V_{ud}$, the Cabibbo-Kobayashi-Maskawa (CKM) flavor mixing between down and up quarks $|V_{ud}| =0.97367 \pm 0.00032$~\cite{ParticleDataGroup:2024cfk}. We further included in~\req{eq:taunqual} the Coulomb correction between electron and proton, proton recoil, nucleon size correction etc. achieved by the additional factor~\cite{Czarnecki:2018okw,
Czarnecki:2004cw,Wilkinson:1982hu} $\mathcal{F}=1.0322$. 

With these inputs \req{eq:taunqual} gives $\tau_n^0=914.6\pm0.6\,(V_{ud})\pm1.5\,(g_A)$\,s. The factor $\mathcal{F}$ does not contain the electroweak radiative correction $1+\mathrm{RC}=1.0390$~\cite{Czarnecki:2004cw,Czarnecki:2018okw}; including it gives $\tau_n^0=4908.6(1.9)\,\mathrm{s}/\big[|V_{ud}|^2(1+3g_A^2)\big]=880.5$\,s~\cite{Czarnecki:2018okw}. In total the theoretical decay rate compares well to the experimental result: The neutron lifetime has been measured with high precision in laboratory experiments using two approaches, the bottle method and the beam method. In the former neutrons are trapped and the surviving neutrons are counted with varying storage times. In the beam method, observed detection rate of the decay products gives the neutron decay rate. The beam method~\cite{Yue:2013qrc} (two experimental results) gives an average lifetime of $\tau_n^\mathrm{beam} = 888.1 \pm 2.0$\,s, while the trap method (8 experimental results since 1990) gives $\tau_n^\mathrm{trap} = 878.36 \pm 0.45$\,s~\cite{Wietfeldt:2024oku}. The Particle Data Group~\cite{ParticleDataGroup:2024cfk} (PDG-2024) presents as a reference value the trap value $\tau_n^0 = 878.4 \pm 0.5$\,s disregarding, as it seems, the beam method measurement.

Although this method-dependence of $\tau_n^0$ may originate from unresolved experimental systematics, we offer a further possibility without claiming any direct evidence: a residual environment dependence of the value of the symmetry breaking Weinberg angle $s^2_\mathrm{W}$ as discussed in Section~\ref{SiW}. 

\subsection{Medium effects in neutron lifespan}\label{DecayMedium}
In this section, we explore medium effects modifying the neutron decay rate in the primordial Universe: The medium effects we retain are the Fermi blocking factors $(1-f_i)$ for all final-state leptons and protons in the in-medium phase-space integrals, the gradual reheating of the photon fluid by $e^+e^-$ annihilation set out in Appendix~A, and the consequent modification of the Hubble rate via $g_*(T)$ in the same epoch. We do not include thermal mass shifts of the electron and nucleons, thermal-photon vertex corrections to the matrix element, or recomputation of the short-distance Coulomb-recoil factor $\mathcal{F}=1.0322$ in medium, nor equally insignificant time dilation originating in particle thermal motion~\cite{Kuznetsova:2010pi}, as in our case, for neutrons with $T/m<10^{-3}$, this effect is negligible. In our evaluation we by default consider noticeable plasma influence on the neutron lifetime in the primordial Universe. Dominant omitted corrections have been already addressed in the precision-BBN literature, in particular by Pitrou--Coc--Uzan--Vangioni~\cite{Pitrou:2018cgg}, and are estimated there at the $\sim 0.1$--$0.5\%$ level on $X_n$ in the temperature range $T \lesssim 1$\,MeV --- well below the variation induced by the percent-level $s^2_\mathrm{W}$ shifts we explore in this work. We do not expect that a full finite-temperature, field-theory based treatment of $n\leftrightarrow p$ rates could influence our results noticeably.
\begin{enumerate}
\item The decay Fermi blocking factors: In the relevant temperature range after neutron freeze-out $T_f\geqslant T\geqslant T_{BBN}$, electrons and neutrinos in the background plasma can reduce the neutron decay rate by Fermi blocking the neutron decay products, this can noticeably lengthen the neutron lifetime compared to its vacuum value. The Pauli blocking has been recognized in the literature~\cite{Wagoner:1966pv,Pitrou:2018cgg,Bernstein:1988ad,Yang:2018qrr}, and the medium dependence of particle decay was recognized by Kuznetsova et al.~\cite{Kuznetsova:2010pi}. We use their method to quantify the ambient cosmic plasma effect on neutron decay, including as a new element the consideration of variation of $s^2_\mathrm{W}$.
\item Gradual reheating by $e^+e^-$-pair annihilation: We obtain a result for $H$ which accounts for disappearing $e^+e^-$-pairs in gradual manner with slowly growing difference in the two different temperatures in cosmic plasma: In our model the already decoupled (frozen-out) neutrinos remain undisturbed. However, as the temperature decreases the thermal abundance of $e^+e^-$-pairs decreases and the entropy from pair annihilation feeds into photons and charged particles, leading to reheating and gradually growing $T>T_\nu$. Gradual reheating impacts the speed of the expansion of the Universe, see Appendix~A, and thus neutron abundance at the edge of BBN epoch.
\end{enumerate}

In the primordial Universe the rate of all present neutron decays per volume in medium, at finite temperature can be written as~\cite{Kuznetsova:2010pi}
\begin{align}\label{eq:ndecayIR}
\frac{dW_{n\to pe\bar{\nu}}}{dVdt}
&= \int\, g_nf_n\frac{d^3p_n}{(2\pi)^3 2E_n}2m_n\bigg[\frac{1}{2m_n}\int\frac{d^3p_{\bar{\nu}}}{(2\pi)^32E_{\bar{\nu}}}\frac{d^3p_p}{(2\pi)^32E_p}\frac{d^3p_e}{(2\pi)^32E_e}\notag\\
&(2\pi)^4\delta^4\left(p_n-p_p-p_e-p_{\bar{\nu}}\right)
\langle|\mathcal{M}|^2\rangle\big[1-f_p\big]\big[1-f_e\big]\big[1-f_{\bar{\nu}}\big]\bigg],
\end{align} 
where the phase-space factors $(1-f_i)$ are Fermi suppression (blocking) factors in the medium. 

Considering the nonrelativistic neutron $E_n\approx m_n$, the thermal neutron decay rate per unit volume, \req{eq:ndecayIR}, can be written as
\begin{align}
\frac{dW_{n\to pe\bar{\nu}}}{dVdt}
=n_n\,\Gamma_{n}.
\end{align} 
This leads to the decay rate of one single neutron in the rest frame of the heat bath, compare~\req{VacNdecay}
\begin{align}
\Gamma_{n}= &\frac{1}{2m_n}\int\frac{d^3p_{\bar{\nu}}}{(2\pi)^32E_{\bar{\nu}}}\frac{d^3p_p}{(2\pi)^32E_p}\frac{d^3p_e}{(2\pi)^32E_e} \notag\\
&
(2\pi)^4\delta^4\left(p_n-p_p-p_e-p_{\bar{\nu}}\right)\langle|\mathcal{M}|^2\rangle
\big[1-f_p\big]\big[1-f_e\big]\big[1-f_{\bar{\nu}}\big]\,.\label{PlasmaNdecay}
\end{align}

To simplify~\req{PlasmaNdecay}: We insert the form of matrix element~\req{VacNdecay} in \req{PlasmaNdecay}; we use for the proton in nonrelativistic limit the three-momentum $d^3p_p$ part of the four-dimensional delta function. The energy-momentum conservation then fixes the antineutrino energy as $E_{\bar\nu}=Q-E_e$ in the neutron rest frame. The remaining $d^3p_{\bar\nu}$ integral is performed in the limit $m_\nu\to 0$, yields a factor $4\pi E_{\bar\nu}^2$, while the $d^3p_e$ integral written in spherical coordinates contributes $4\pi p_e E_e\,dE_e$ with $p_e=\sqrt{E_e^2-m_e^2}$. Collecting these factors and imposing the kinematic bounds $m_e\leq E_e\leq Q$, one recovers the standard vacuum phase-space integral for the neutron decay multiplied by the in-medium Fermi distribution blocking factors $1-f_i$ for electron, proton, and antineutrino. We keep for now the blocking fugacity $\Upsilon_\nu$ for the (free-streaming~\cite{Birrell:2012gg}) neutrino background. Separating out the reaction strength $D$,~\req{DimFac}, the reduced decay rate written using dimensionless variable $\xi=E_e/Q$ is
\begin{align}
\label{Decay_rate_01}
 \Gamma_{n}^\prime= 
 \int^1_{m_e/Q}\!\!\!\! d\xi\,\frac{\xi(1-\xi)^2}{\displaystyle e^{-Q\xi/{T}}+1}\,\frac{\sqrt{\xi^2-(m_e/Q)^2}}{\Upsilon_\nu\displaystyle e^{-Q(1-\xi)/T_\nu}+1}\,. 
\end{align} 

This shows how the primordial plasma of the Universe modifies the neutron decay phase space in terms of both the photon temperature $T$ and the neutrino effective temperature $T_\nu$, which is presented in Appendix~A at the required level of precision. $Q$ is the neutron-proton mass difference seen in~\reqs{eq:Qmdif}{eq:QmdifEx}. The limit of the decay rate in vacuum is found in~\req{Decay_rate_01} ignoring the Fermi function denominators. The use of free-streaming neutrino distribution characterized by the non-reheated temperature $T_\nu$ is justified by the fact that neutrino kinetic decoupling~\cite{Birrell:2014uka,Mangano:2005cc,Akita:2020szl} occurs at $T_k \simeq 1.5$\,MeV, well above the temperature range $T \lesssim 1$\,MeV where the in-medium correction to $\tau_n$ is computed. The residual non-equilibrium corrections to the neutrino phase-space distribution following decoupling have been quantified in our earlier work~\cite{Birrell:2014uka,Birrell:2012gg} and reviewed in detail in~\cite{Rafelski:2024fej}; they are well below the precision required for the present calculation.

We do not expect that the short distance Coulomb correction between electron and proton, the proton recoil, nucleon size correction etc. are modified in the cosmic plasma. Thus we adapt the factor $\mathcal{F}$ into our calculation at finite $T$ and use the following neutron decay rate in the cosmic plasma 
\begin{align}
\label{Decay_rate_02}
\Gamma_n=D\mathcal{F}\,\Gamma^\prime_n
\end{align}

\begin{figure}
\begin{center}
\hspace*{-0.6cm} \includegraphics[width=5.2in]{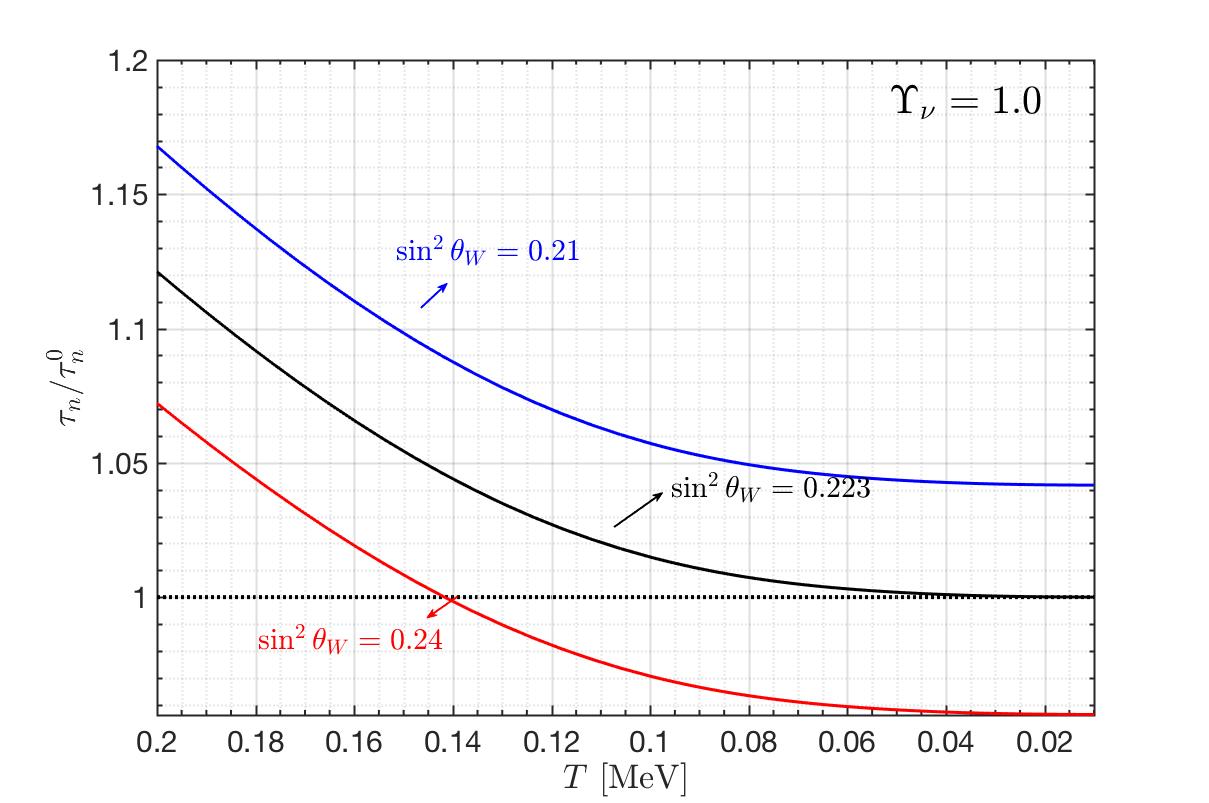}
\caption{Neutron lifetime with respect to vacuum reference value $\tau_n/\tau_n^0$ as a function of plasma temperature $T$ for three different values of the Weinberg angle $s^2_\mathrm{W}$ with range $0.21, 0.223, 0.24$. The displayed $s^2_\mathrm{W}$ range brackets the current experimental determinations: $s^2_\mathrm{W}(m_Z) = 0.2313$ at the $Z$ scale, the global low-energy value near $0.238$, and the CODATA recommended value $0.223$ adopted in this work.}
\label{LifeWeinberGFig}
\end{center}
\end{figure}

In Fig.~\ref{LifeWeinberGFig}, we show the ratio of the neutron lifetime in a cosmic plasma, $\tau_n$, to its vacuum value, $\tau_n^0$, as a function of plasma temperature $T$ for three different values of the Weinberg angle, $s^2_\mathrm{W}$. The values shown bracket the spread of current experimental determinations of $s^2_\mathrm{W}$. The curves indicate that at high temperatures, the neutron lifetime increases significantly relative to vacuum value due to Fermi blocking effects from the surrounding background. As the temperature decreases towards edge of BBN epoch, the ratio approaches a constant value, converging to the respective vacuum lifetime. Larger values of $s^2_\mathrm{W}$ generally correspond to longer neutron lifetimes, while smaller values lead to shorter lifetimes. Within the empirically motivated $s^2_\mathrm{W}$ range shown, the modification of $\tau_n$ exceeds the beam-bottle discrepancy by a significant factor. This result demonstrates that even a modest variation in the electro-weak mixing angle can substantially influence neutron decay rates in the primordial plasma Universe conditions, exceeding the effect of Pauli blocking. 

We also note the zero-temperature limits, corresponding to the vacuum lifespan studied in Section~\ref{nlifetime}: the near one-percent difference between beam and bottle methods is significantly smaller than the spread shown above. We question, without claiming any direct evidence, the possibility that some part of the laboratory $\tau_n$ scatter could arise from a small environmental variation of $s^2_\mathrm{W}$. This is not as an established conclusion but a motivation for further study.

\subsection{In plasma proton to neutron conversion}\label{ProtonDecayMedium}
In the primordial plasma there is also the inverse to neutron decay back-reaction $ pe\bar{\nu}\to n$. Intuitively one expects this process to be relevant when the temperature is much larger compared to the mass difference between proton and neutron, and negligible in the main temperature domain of interest to us. For this reason, the following is a sideline and can be omitted on first read of this work.

Replacing in \req{eq:ndecayIR} Fermi blocking factor $1-f_i$ by abundance factor $f_i$ and conversely $f_n\to 1-f_n$, we obtain the inverse reaction rate per unit volume
\begin{align}\label{eq:nprodIR}
\frac{dW_{pe\bar{\nu}\to n}}{dVdt}
&= \int\, [1-f_n]\frac{d^3p_n}{(2\pi)^3 2E_n} \int\frac{d^3p_{\bar{\nu}}}{(2\pi)^32E_{\bar{\nu}}}\frac{d^3p_p}{(2\pi)^32E_p}\frac{d^3p_e}{(2\pi)^32E_e}\notag\\
&(2\pi)^4\delta^4\left(p_n-p_p-p_e-p_{\bar{\nu}}\right)
\langle|\mathcal{M}|^2\rangle g_pf_p g_ef_e g_{\bar{\nu}}f_{\bar{\nu}}.
\end{align} 
\indent As implied by \req{eq:nprodIR}, thermal electrons and antineutrinos hit a nearly resting proton to produce a neutron which is escaping at any energy and momentum that results, consistent with the conservation law of energy and momentum, and accounting for the energy needed to convert a proton into neutron. The meaning of nonrelativistic approximation is now not only to set $E_p=m_p$. In addition, when going to the rest frame of the proton which due to its high mass is hardly moving, in the original laboratory rest frame we keep unchanged the statistical distribution of relativistic electrons and neutrinos.

In the phase space integral \req{eq:nprodIR} we set in the nonrelativistic approximation the denominator $2E_p\to 2m_p$ and we keep the statistical distributions in the rest frame of laboratory which allows the multiplicative separation 
\begin{align}
\frac{dW_{pe\bar{\nu}\to n}}{dVdt}
=n_p\,\Gamma_{p}\,,
\end{align} 
where we use the subscript $p$ for $pe\bar{\nu}\to n$ process
\begin{align}\label{eq:nprodG}
\Gamma_{p}= & 
 \frac{1}{2m_p}\int\, \frac{d^3p_n}{(2\pi)^3 2E_n}\frac{d^3p_{\bar{\nu}}}{(2\pi)^32E_{\bar{\nu}}}\frac{d^3p_e}{(2\pi)^32E_e}\notag\\&
 (2\pi)^4\delta^4\left(p_n-p_p-p_e-p_{\bar{\nu}}\right)
\langle|\mathcal{M}|^2\rangle (1-f_n) g_e\,f_e g_{\bar{\nu}}f_{\bar{\nu}}\,.
\end{align} 
\indent The formulation presented allows us to view the rate $\Gamma_{p}$ as the induced `decay rate' of the proton into neutron in plasma. The similarity of this result with \req{PlasmaNdecay} allows us to separately study the phase space part removing the intrinsic rate factor $D$ \req{DimFac}, evaluating
\begin{align}\label{eq:pprodredG}
\Gamma^\prime_p=\Gamma_p/D=\int_{m_e/Q}^{1} d\xi\;
\frac{\xi(1-\xi)^2\sqrt{\xi^2-({m_e}/{Q})^2}} {\bigl[e^{Q\xi/T}+1\bigr]\,\bigl[\Upsilon_\nu^{-1}\,e^{Q(1-\xi)/T_\nu}+1\bigr]}.
\end{align}
We have omitted the Fermi blocking factor $(1-f_n)\approx1$ to $10^{-13}$ precision. The range of the integral is the same as in \req{Decay_rate_01}; in the argument of the integral the value of $Q$ is reversed in sign since this energy is needed rather than gained. Similarly any excess of neutrinos described by $\Upsilon>1$ would not block but lead to gain so the inverse of the power seen in \req{Decay_rate_01} appears in \req{eq:pprodredG}.

To evaluate the relation between the forward and backward rates we cast \req{eq:pprodredG} into a form similar to \req{Decay_rate_01} by multiplying the nominator and denominator of the two Fermi distribution factors by, respectively, $e^{-Q\xi/T}$ and $\Upsilon_\nu \,e^{-Q(1-\xi)/T_\nu}$, yielding the same expressions in denominators for both equations \req{Decay_rate_01} and \req{eq:pprodredG} with a multiplicative factor (product of above factors) in the nominator of \req{eq:pprodredG}. We than note that when $T_\nu \to T$, thus before reheating due to $e^+e^-$-annihilation we have the detailed balance relation 
\begin{align}
\Gamma_p=\Upsilon_\nu e^{-Q/T}\Gamma_n\,,\qquad
\frac{dW_{pe\bar{\nu}\to n}}{dVdt}=\Upsilon_\nu e^{-Q/T}\frac{dW_{n\to pe\bar{\nu}}}{dVdt}\,, \qquad T_\nu \to T \,.
\end{align}
The effect of reheating $T>T_\nu$ is to make the relatively small ratio $\Gamma_p/\Gamma_n= W_{pe\bar{\nu}\to n}/W_{n\to pe\bar{\nu}}$ slightly larger for $T<0.5$\,MeV. This small breach of detailed balance originating in the non-thermal (free-streaming) neutrino background is, however, of no consequence: in this temperature domain the two-body conversion reactions $n\nu \leftrightarrow pe$ and $ne^{+} \leftrightarrow p\bar{\nu}$ dominate $\Gamma_p$ by many orders of magnitude, as is visible in Fig.~\ref{RelaxationRate_fig} which we discuss 
in Section~\ref{WInpRates} below. At low temperature detailed balance is also very slightly broken by neutron decay competing with the dominant two-body rates; this small effect is what produces the rising ratio $X_n^{th}/X_n^{ad}$ visible in the insert of Fig.~\ref{Xn001fig} which we discuss 
in Section~\ref{Neutron} below. 

\section{Primordial Neutron Concentration}\label{Abundance}
\subsection{Thermal model}\label{ThermalM} 
In the primordial Universe we study how weak reaction rates impact the neutron abundance with interest to understand precisely their abundance at onset of BBN. We first look at a naive description: Assuming thermal equilibrium yields of particles, we define the neutron concentration $X_n^f$ at instantaneous in temperature weak interaction (WI) freeze-out:
\begin{align}
\label{Xnabundance2}
X_n^f \equiv \frac{n_n^f}{n_n^f+n_p^f}= \frac{1}{1+\displaystyle e^{Q/T_f}},
\end{align}
where $n_n^f=n_n(T_f)$ and $n_p^f=n_p(T_f)$ are freeze-out neutron and proton thermal densities, respectively, obtained at the freeze-out temperature $T_f$. $X_n^f$ depends on temperature $T_f$ at which neutrons decouple from the heat bath, and the neutron-proton mass difference (in medium) $Q$. 

Following freeze-out, the free-streaming neutron concentration is subject only to decay,
\begin{align}
\label{Xn_abundance}
X_n(T_{BBN})=X_n^f\exp\left(-\frac{t_{BBN}-t_f}{\tau_n}\right)\,.
\end{align}
We see that the neutron concentration $X_n(T_{BBN})$ provides an important probe of the underlying WI reaction rates governing the freeze-out condition, and to some extent a probe of neutron lifetime ${\tau_n }$ in the primordial plasma. Normally the neutron lifetime in the vacuum, $\tau_n\to \tau_n^0$, is used to calculate the neutron concentration and the inputs are finessed to obtain the \lq desired\rq\ value $X_n(T_{BBN})\approx0.13$. Many textbook presentations of BBN argue in this manner; indeed, a recent study of how neutron lifespan influences BBN also took this line of thought~\cite{Chowdhury:2022ahn}, which was also present in our earlier study~\cite{Yang:2018qrr}. 

The above considerations require the empirical input of neutron freeze-out condition and treats subsequent evolution without consideration of background scattering effects, thus provides a qualitative estimate of neutron concentration at best. The following full kinetic model study shows that these qualitative arguments do not suffice.

\subsection{WI Neutron and Proton Reaction Rates in the Primordial Plasma}\label{WInpRates}
We now show how the neutron abundance is governed by the competition between WI reaction rates governing conversion reactions between neutrons and protons, and the Hubble expansion rate. Since the Fermi constant $G_\mathrm{F}$ governing these rates depends on the Weinberg angle $s^2_\mathrm{W}$, variations in this angle modify weak rates while the Hubble parameter remains set by the plasma energy density. This is the mechanism causing the neutron abundance to depend on $s^2_\mathrm{W}$ and thus to impact the BBN. 

The neutron abundance is impacted by the weak interaction reactions between neutron and proton via the dense thermal cosmic neutrino background: 
\begin{align}
 &n+\nu_e\longleftrightarrow p+e^-,\\
 &n+e^+\longleftrightarrow p+\bar{\nu}_e.
\end{align}
These are charged current reaction matrix elements mediated by the W-gauge Boson akin to the neutron decay. Therefore transition amplitudes for these nuclear reactions can be written just like those seen in the neutron decay paying attention to the momentum four-vectors of the respective reacting particles:
 \begin{align}
 &|M_{n\nu\to pe}|^2= G_\mathrm{F}^2V_{ud}^2\left(1+3g^2_A\right)16(p_n\cdot p_\nu)(p_p\cdot p_e)\,,\\
 &|M_{ne\to p\nu}|^2=G_\mathrm{F}^2V_{ud}^2\left(1+3g^2_A\right)16(p_n\cdot p_{e^+})(p_p\cdot p_{\bar{\nu}})\,.
 \end{align}
Considering the nonrelativistic limit for neutron and proton, the transition matrix becomes:
\begin{align}
&|M_{n\nu\to pe}|^2=G_\mathrm{F}^2V_{ud}^2\left(1+3g^2_A\right)16\left(m_nE_\nu\right)\left(m_pE_e\right),\\
&|M_{ne\to p\nu}|^2= G_\mathrm{F}^2V_{ud}^2\left(1+3g^2_A\right)16(m_nE_e)(m_pE_\nu).
\end{align}

The Lorentz invariant transition probability per unit time and unit volume corresponding to the process $1+2\to3+4$ is
\begin{align}
\frac{dW_{12\rightarrow34}}{dVdt} = \frac{g_1g_2}{1+I} &\int\frac{d^3\mathbf{p}_1}{(2\pi)^3\,2E_1}\frac{d^3\mathbf{p}_2}{(2\pi)^3\,2E_2}\frac{d^3\mathbf{p}_3}{(2\pi)^3\,2E_3} \frac{d^3\mathbf{p}_4}{(2\pi)^3\,2E_4}\notag\\
&
(2\pi)^4\delta^{(4)}(p_1+p_2-p_3-p_4) |M_{12\leftrightarrow34}|^2\,f_1\,f_2\,(1- f_3)\,(1-f_4),
\end{align}
where the phase-space factors $(1-f_i)$ are Fermi suppression factors in the medium, and the symmetry factor $1/(1+I)$ with $I=0$ for distinguishable particles in the initial states or final states, and $I=1$ for indistinguishable particles.

Substituting the thermal (Fermi) distribution function for the electron, the free streaming neutrino~\cite{Birrell:2012gg}, nonrelativistic proton and neutron, and integrating over the delta function, the thermal reaction rate per volume takes the form
\begin{align}
&\frac{dW_{n\nu\rightarrow pe}}{dVdt}=n_n\,\Gamma_{n\nu\to pe},\qquad\qquad \frac{dW_{ne\rightarrow p\nu}}{dVdt}=n_n\,\Gamma_{ne\to p\nu}\,,
\end{align}
where the thermal reaction rate $\Gamma$ integrated in terms of dimensionless variables $q={E_\nu}/{Q}+1$ and $\eta=E_e/Q$, respectively, are 
\begin{align}
\Gamma_{n\nu\to pe}\equiv D\Gamma^\prime_{n\nu\to pe}\,,
\qquad 
&\Gamma^\prime_{n\nu\to pe} = \int^\infty_{1} dq\frac{q(q-1)^2\sqrt{q^2-(m_e/Q)^2}}{\left[1+\Upsilon_\nu^{-1}\displaystyle e^{{Q(q-1)}/{T_\nu}}\right]\left[1+\displaystyle e^{-Qq/T}\right]}\,,\\[0.4cm]
\Gamma_{ne\to p\nu}\equiv D\Gamma^\prime_{ne\to p\nu}\,,
\qquad 
&\Gamma^\prime_{ne\to p\nu} = \int_{m_e/Q}^{\infty} d\eta\frac{\eta(\eta+1)^2\sqrt{\eta^2-(m_e/Q)^2}}{\left[1+\Upsilon_\nu\displaystyle e^{-{Q(\eta+1)}/{T_\nu}}\right]\left[1+\displaystyle e^{{Q\eta}/{T}}\right]}\,.
\end{align}

For the reverse reaction, the Lorentz invariant transition probability per unit time and unit volume corresponding to the given reactions can also be written as above
\begin{align}
&\frac{dW_{pe\rightarrow n\nu}}{dVdt}=n_p\,\Gamma_{pe\to n\nu},\qquad\qquad\frac{dW_{p\nu\rightarrow ne}}{dVdt}=n_p\,\Gamma_{p\nu\to ne}\,,
\end{align}
where the reaction rates are given by
\begin{align}
 \Gamma_{pe\to n\nu }\equiv D\Gamma^\prime_{pe\to n\nu }\,,
 \qquad 
&\Gamma^\prime_{pe\to n\nu }=\int^\infty_{1} dq\frac{q(q-1)^2\sqrt{q^2-{(m_e/Q)^2}}}{\left[1 + \Upsilon_\nu\displaystyle e^{{-Q(q-1)/T_\nu}}\right]\left[1 + \displaystyle e^{{Qq/T}}\right]}\,,
\\[0.4cm]
 \Gamma_{p\nu\to ne}\equiv D\Gamma^\prime_{p\nu\to ne}\,,
 \qquad
 &\Gamma^\prime_{p\nu\to ne}=\int_{m_e/Q}^{\infty} d\eta\frac{\eta(\eta+1)^2\sqrt{\eta^2-(m_e/Q)^2}}{\left[1+\Upsilon_\nu^{-1}\displaystyle e^{ {Q(\eta+1)}/{T_\nu}}\right] \left[1+\displaystyle e^{ {-Q\eta}/{T}}\right]}\,\,.
\end{align}

\begin{figure}[t]
\begin{center}
\hspace*{-0.5cm} \includegraphics[width=5in]{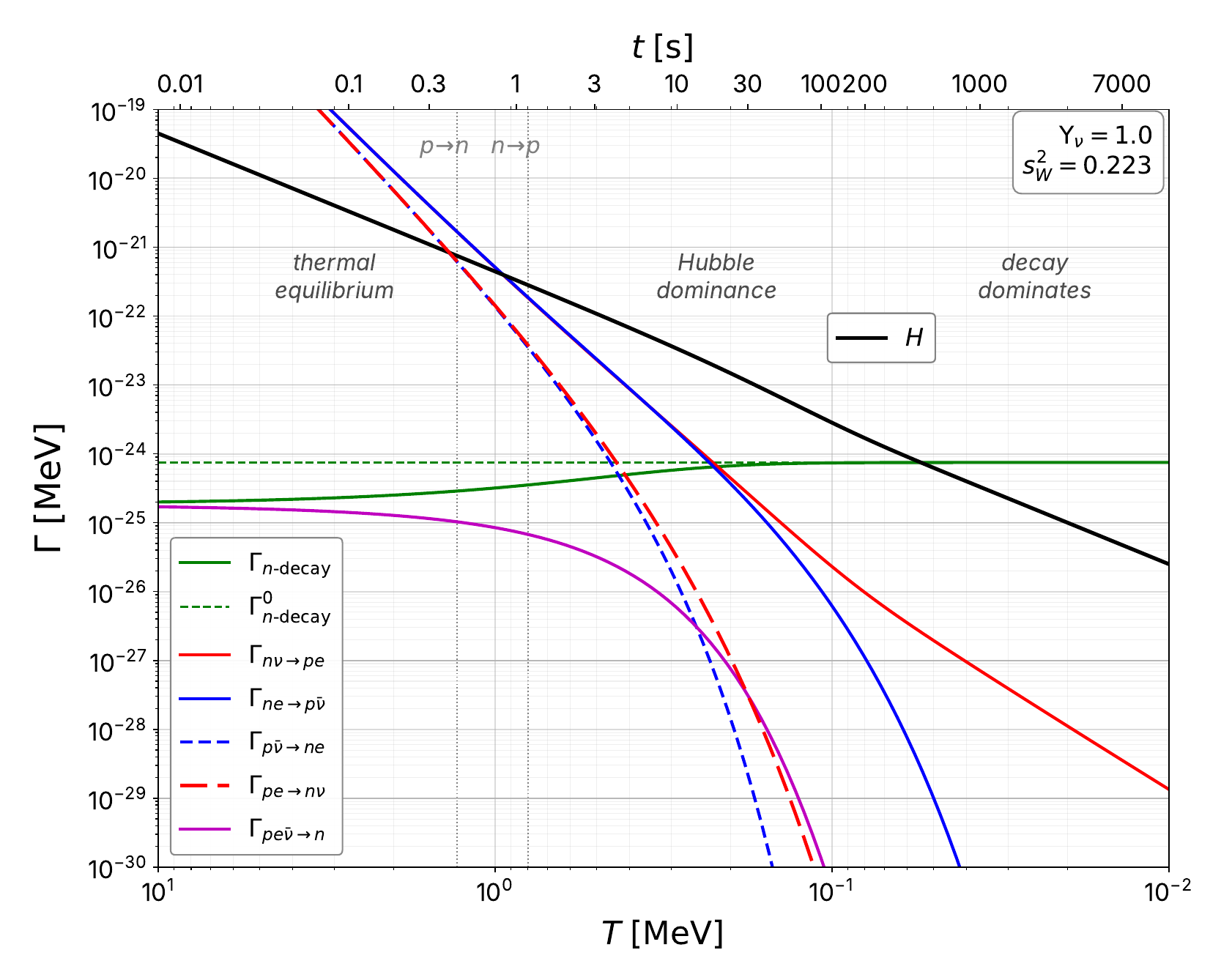}
\caption{Thermal reaction rates as a function of temperature $10\geqslant T\geqslant0.01$MeV for processes shown in subscripts. The top axis shows cosmological time $t$\,[s] anchored at~\cite{Rafelski:2024fej} $t(150\,\mathrm{MeV})=25\,\mu\mathrm{s}$, with $\Upsilon_\nu=1.0, s^2_\mathrm{W}=0.223$.}
\label{RelaxationRate_fig}
\end{center}
\end{figure}
In Fig.~\ref{RelaxationRate_fig} we show the thermal reaction rates as a function of temperature $T\geqslant0.01$\,MeV for normal strength of WI with Weinberg angle $\sin^2\theta_\mathrm{W}=0.223$. We observe the following
\begin{itemize}
\item \textbf{Thermal equilibrium epoch:}\\
When $T>1$\,MeV, weak interaction rates between neutron and proton dominate Hubble rate. Since expansion of the Universe is \lq slow\rq\ the neutron and proton abundances are in thermal equilibrium. Looking closer we see that the neutron-to-proton conversion decouples (becomes smaller compared to Hubble expansion) at temperature $T\simeq 0.8$\,MeV and the proton-to-neutron conversion decouples at $T\simeq 1.3$\,MeV. The magenta curve $\Gamma_{pe\bar\nu\to n}$ shows the three-body back-reaction rate~\req{eq:nprodG} can be seen well below all other rates.
\item \textbf{Hubble dominance kinetic epoch:}\\
Below above boundaries, in the range $1.0>T>0.05$\,MeV, the dominant rate is the Hubble expansion rate. In this temperature window we could naively expect that the neutron abundance only dilutes due to Universe expansion considering that all conversion reactions are smaller than the Hubble expansion rate $H$. In this regime, attention must be paid to the exact value of $H$ as we encounter the transition between Hubble parameter being determined by energy density contributions of photons, neutrinos, and electrons/positrons to contributions from only photons and neutrinos driven by the smooth disappearance of $e^\pm$-pairs.
\item \textbf{Neutron decay dominated epoch:}\\
When $0.05>T>0.01$\,MeV (and below), the dominant rate is the neutron decay rate, the neutron abundance is decreased by neutron decay in the primordial Universe.
\end{itemize}
In summary, there is a well-defined sequence of dominant here relevant reaction rates as the Universe cools. To accurately understand the evolution of neutron abundance in the primordial Universe, it is essential to analyze the corresponding kinetic evolution equations in detail.
 
\subsection{Kinetic Neutron Population Equation}\label{Neutron}
We now develop a kinetic model for the neutron concentration, i.e. the abundance ratio between neutrons and baryons, 
\begin{align}
X_n=\frac{N_n}{N_B}=\frac{N_n}{N_n+N_p}\,,
\end{align}
where $N_n$ and $N_p$ are the \lq comoving\rq\ in the expanding volume numbers of neutron and protons respectively. The kinetic equation for the concentration of neutron can be written as
\begin{align}\label{dXndt}
\frac{dX_n}{dt}&=\frac{d}{dt}\!\left(\frac{N_n}{N_B}\right)=\frac{1}{n_B}\left(\frac{1}{V}\frac{dN_n}{dt}\right)\,,
\end{align}
where we consider the comoving number of baryon is constant after the baryon genesis epoch; comoving means that we study the abundance in cosmologically expanding volumes.

The population of neutrons is governed by gains and losses due to reactions we considered 
\begin{align}
 \frac{1}{V}\frac{dN_n}{dt}=\frac{dW_{pe\rightarrow n\nu}}{dVdt}+\frac{dW_{p\nu\rightarrow ne}}{dVdt}-\frac{dW_{n\nu\rightarrow pe}}{dVdt}-\frac{dW_{ne\rightarrow p\nu}}{dVdt}-\frac{dW_{n\rightarrow pe\nu}}{dVdt}\,.
\end{align}
Substituting the Lorentz invariant transition probability per unit time and unit volume corresponding to the given reactions, the population equation of neutrons becomes:
\begin{align}
 \frac{1}{V}\frac{dN_n}{dt}= \,\,n_p\left(\Gamma_{pe\rightarrow n\nu}+\Gamma_{p\nu\rightarrow ne}\right) -n_n\left(\Gamma_{n\nu\rightarrow pe}+\Gamma_{ne\rightarrow p\nu}\right)-n_n\,\Gamma_{n}+n_p\Gamma_p\,.
\end{align}
Inserting into~\req{dXndt} we obtain the kinetic evolution of neutron concentration.

It is convenient to introduce now the following relaxation rates:
\begin{align}
 \Gamma_{p\to n}=\Gamma_{pe\rightarrow n\nu}+\Gamma_{p\nu\rightarrow ne}\,, 
 \Gamma_{n\to p}=\Gamma_{n\nu\rightarrow pe}+\Gamma_{ne\rightarrow p\nu}\,,
\end{align}
Then the kinetic equation for the neutron concentration $X_n$ can be written as
\begin{align}
 \frac{dX_n}{dt}&=X_p(\Gamma_{p\to n}+\Gamma_p)-X_n\left(\Gamma_{n\to p}+\Gamma_{n}\right)\,,\notag\\[0.1cm]
 &=\left(1-X_n\right)(\Gamma_{p\to n}+\Gamma_p)-X_n\left(\Gamma_{n\to p}+\Gamma_{n}\right)\,,\notag\\[0.1cm]
 &=\Gamma_{p\to n}+\Gamma_p-X_n\left(\Gamma_{p\to n}+\Gamma_{n\to p}+\Gamma_{n}+\Gamma_p\right)\,.
 \label{KineticXn}
\end{align}

The first approximation we can make to solve~\req{KineticXn} by assuming that the Universe is expanding very slowly and there can be instantaneous kinetic detailed balance equilibrium in the particle yields; this is the \lq adiabatic\rq\ approximate solution found setting $dX_n/dt=0$ in~\req{KineticXn}; it applies in the domain we called thermal equilibrium above. If neutrons would not decay $X^{ad}_n$ would indeed be exactly the thermal result~\req{Xnabundance2}. However, solving with decay rate we obtain the instantaneous detailed balance result
\begin{align}\label{AdSol}
 X^{ad}_n=\frac{1}{1+\Gamma_{n\to p}/(\Gamma_{p\to n}+\Gamma_p)+\Gamma_{n}/(\Gamma_{p\to n}+\Gamma_p)}.
\end{align}
The adiabatic solution arises from the competition between neutron production, weak interaction conversion, and decay rates, yielding the modified equilibrium neutron concentration. 

\begin{figure}[t]
\begin{center}
\hspace*{-0.5cm} \includegraphics[width=5in]{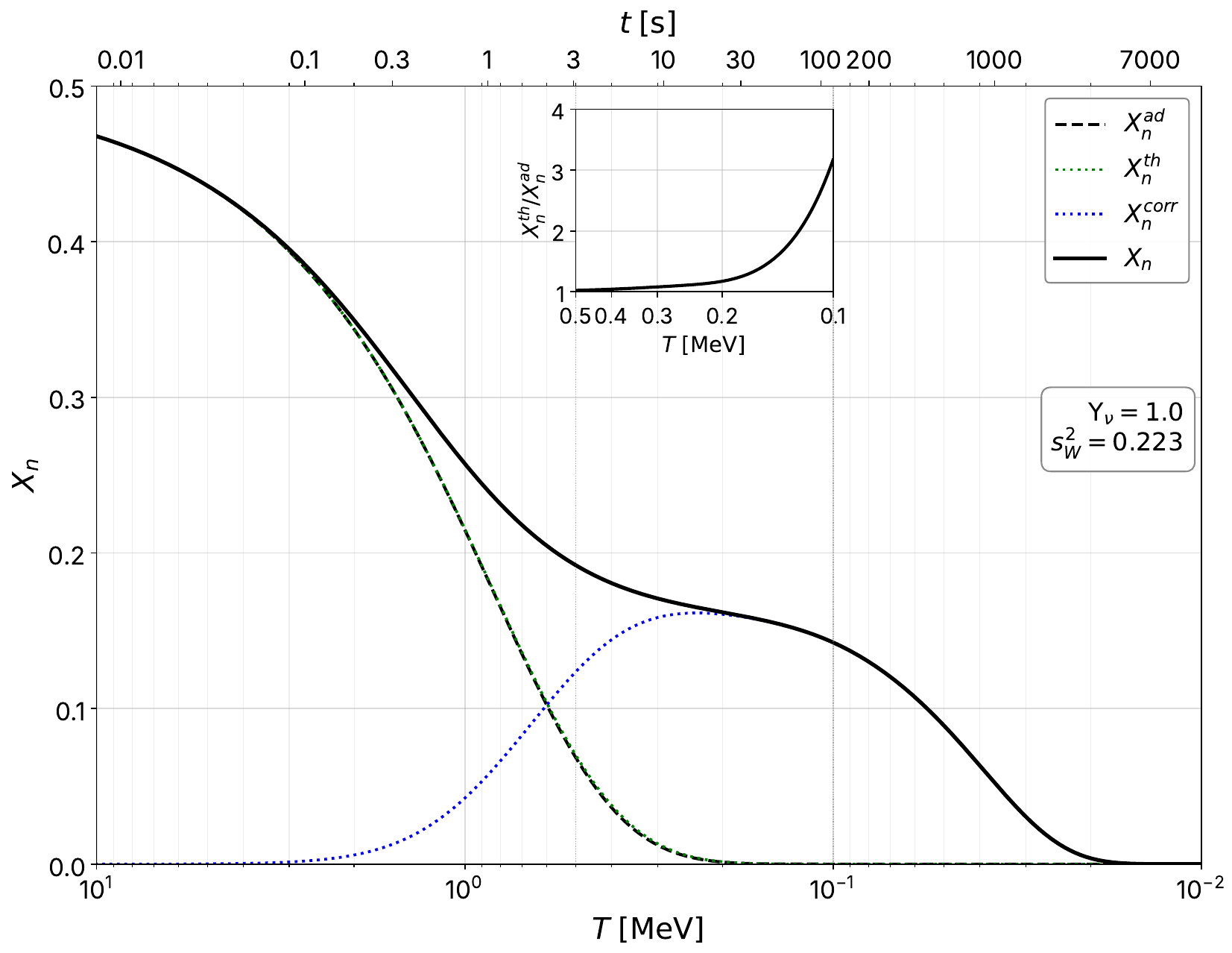}
\caption{The neutron concentration $X_n$ as a function of temperature. Dashed line: Result ($X_n^{ad}$) obtained under assumption that WI dominate Hubble expansion. Green dotted line: Thermal equilibrium $X_n^{th}$, it is hard to see a difference but in the ratio $X^{th}_n/X^{ad}_n$ seen in the insert for $T\in[0.1,0.5]$\,MeV. Blue dotted line: Kinetic correction $X_n^{corr}=X_n-X_n^{ad}$ arising from competition of kinetic neutron abundance processes with Hubble expansion. Solid line: The concentration of neutrons in the Universe $X_n=X_n^{ad}+X_n^{corr}$. The top axis shows cosmological time $t$\,[s] anchored at $t(150\,\mathrm{MeV})= 25\,\mu\mathrm{s}$~\cite{Rafelski:2024fej}. Results obtained for $\Upsilon_\nu=1.0$, $s^2_\mathrm{W}=0.223$ for chemically equilibrated (free-streaming) neutrino background~\cite{Birrell:2012gg} with decoupling at and above $T_\nu=1.5$\,MeV~\cite{Rafelski:2024fej,Birrell:2014uka}.}
\label{Xn001fig}
\end{center}
\end{figure}

It is important to note that in the ratios of $\Gamma$s seen in~\req{AdSol} the strength of weak interactions cancels out exactly, the adiabatic result just like the thermal equilibrium result depends on phase space size of different processes only. This solution $X_n^{ad}$ is shown as dashed line in Fig.~\ref{Xn001fig}, and on linear scale it agrees with the thermal equilibrium result $X_n^{th}$ also shown as overlay green dotted line. Any deviation could be only expected in low temperature domain, we see this in the insert in Fig.~\ref{Xn001fig} showing $X^{th}_n/X^{ad}_n$.
Looking in depth at reaction rates in Fig.~\ref{RelaxationRate_fig} we recognize the cause of this tiny detailed balance breaking to be: i) The reheating which did not alter free-streaming neutrinos, and ii) the presence of neutron decay since the back reaction $\Gamma_p$ is already shut-off due to energy threshold. These effects are small and, as it seems, compensatory.

We now obtain the correction to the adiabatic neutron concentration $X_n^{ad}$. The kinetic equation of neutron concentration can be used in the format
\begin{align}
 \frac{dX_n}{dt}+\left(\Gamma_{p\to n} + \Gamma_{n\to p} + \Gamma_{n}+\Gamma_{p}\right)\left(X_n-X_n^{ad}\right)=0\,.
\end{align}
Adopting the solution method from paper~\cite{Bernstein:1988ad}, the general solution can be written 
\begin{align}\label{Eq:KineticN}
 X_n(y)&=X_n^{ad}(y)-\int^y_{y_0}\,dy^\prime\,I(y,y^\prime)\,\frac{d\,X_n^{ad}(y^\prime)}{dy^\prime}\,,\notag\\[0.2cm]
 &=X_n^{ad}(y)+X_n^{corr}(y)\,,\qquad y\equiv Q/T\,,
\end{align}
where the function $I(y,y^\prime)$ is 
\begin{align}\label{RateRatio}
I(y,y^\prime) = \exp {\bigg[ - \int^y_{y^\prime} \frac{dy^{\prime\prime}}{y^{\prime\prime}}\left(1-\frac{1}{3}\frac{d\ln g^s_\ast}{d\ln y^{\prime\prime}}\right)\frac{\left(\Gamma_{p\to n} + \Gamma_{n\to p} + \Gamma_{n}+\Gamma_{p}\right)}{H}\bigg]}\,.
\end{align}
The Hubble expansion rate competes with the WI reaction rate in the function $I(y,y^\prime)$. The factor in round brackets converts $dy/y$ into $H\,dt$ for isentropic expansion while $e^\pm$ annihilate, see~\req{dtdT} in Appendix~A; for constant $g^s_\ast$ it reduces to the form used in Ref.~\cite{Bernstein:1988ad}. The sign of the correction term is opposite to that in~\req{dtdT} because~\req{RateRatio} is written in the inverse-temperature variable $y^{\prime\prime}=Q/T^{\prime\prime}$. This correction is not negligible: Consider the deuteron abundance peak when neutron capture into deuterons becomes as fast as deuteron burn near to temperature $T\simeq75$\,keV. At this condition this factor lengthens the time from 194\,s to 228\,s and lowers $X_n$ there by about 6\%.

The correction term~\req{Eq:KineticN} is seen as a dotted line in Fig.~\ref{Xn001fig}. The combined result, the neutron concentration as a function of temperature in primordial Universe, is the solid line. The neutron concentration follows the adiabatic solution for high temperatures where WI reactions remain dominant. As the Hubble expansion rate competes with WI reaction rates, the neutron concentration is dominated by kinetic processes only. In between there is a domain of transition. Towards the end of the BBN temperature range $X_n$ decreases toward zero since the neutron decay becomes the dominant process. 

\begin{figure}
\begin{center}
\hspace*{-0.4cm} \includegraphics[width=4.8in]{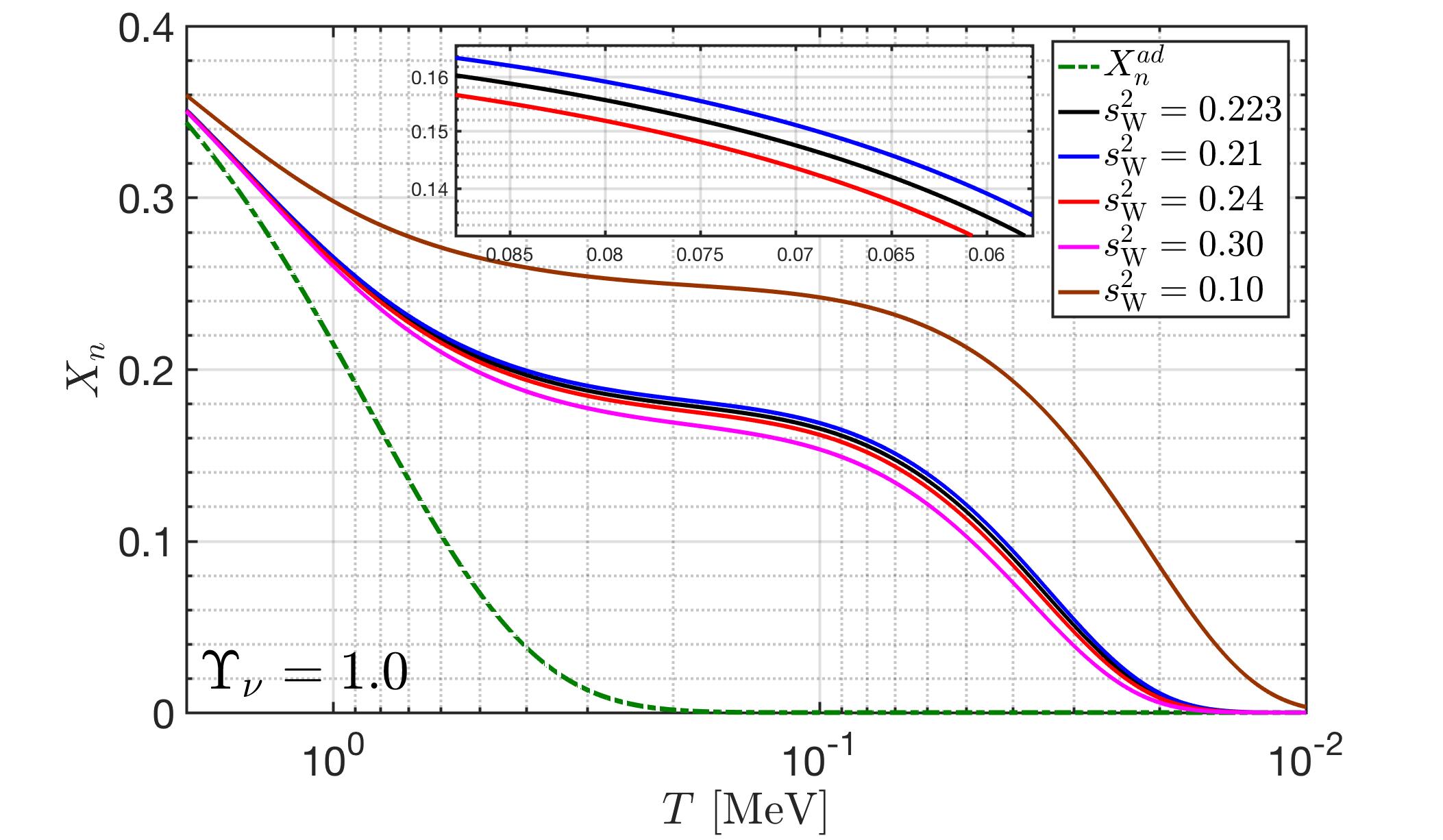}\\
\hspace*{-0.4cm} \includegraphics[width=4.8in]{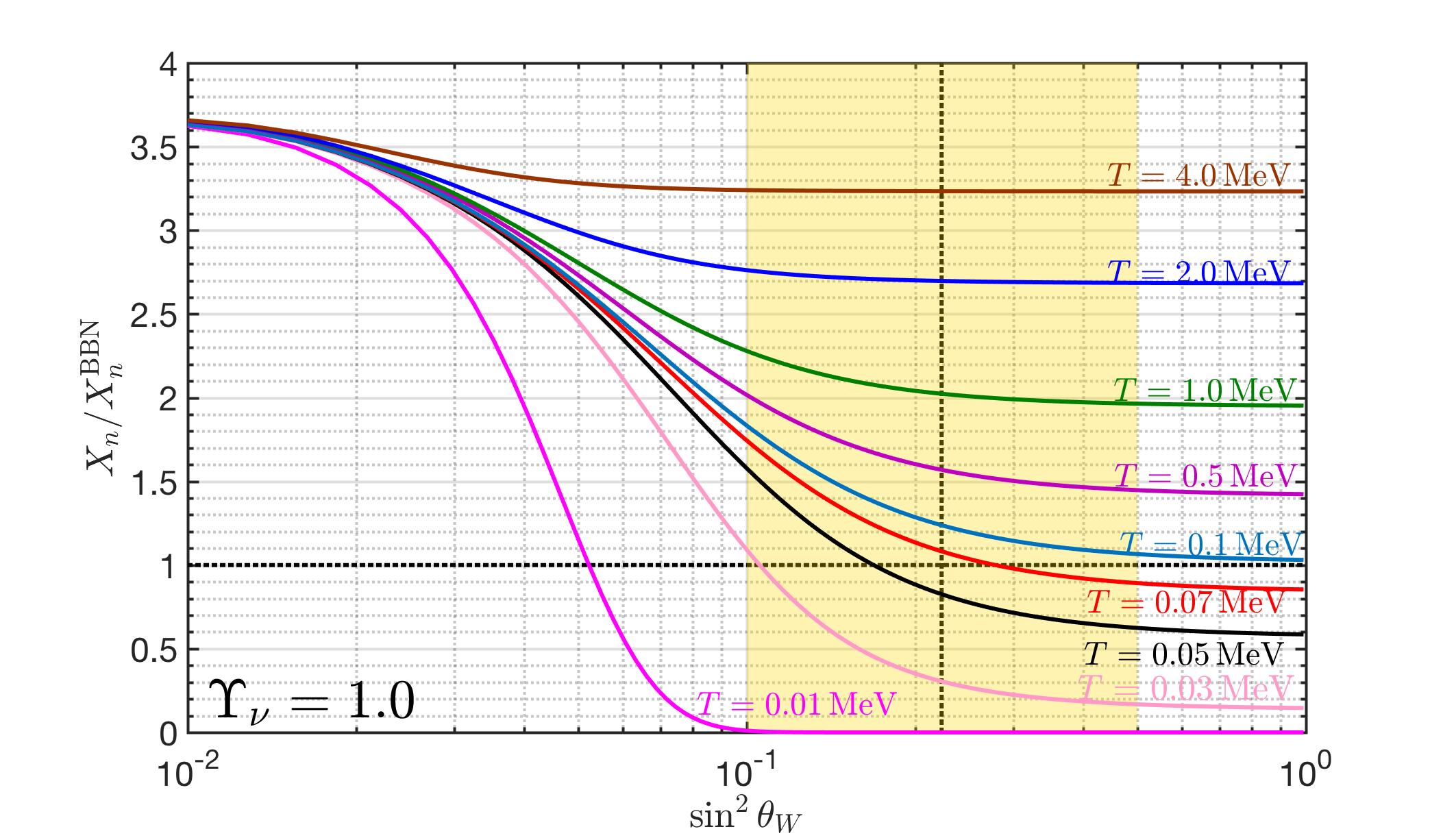}
\caption{\textbf{Top frame:} The neutron concentration $X_n$ as a function of temperature $T$ with different values of Weinberg angle $s^2_\mathrm{W}$ and assumed chemical equilibrium of background neutrinos, $\Upsilon_\nu=1$. The cases $s^2_\mathrm{W}=0.1$ and $s^2_\mathrm{W}=0.3$ illustrate the range of the Weinberg angle for which the approximation in Eq.~(\ref{GFcorr}) remains valid. The insert amplifies variation of neutron concentration for temperature characteristic of temperature at BBN onset with a small variation of $s^2_\mathrm{W}$ from standard value as we considered before. There is no dependence on $s^2_\mathrm{W}$ in the adiabatic component (dashed line); \textbf{Bottom frame:} The normalized neutron concentration $X_n/X_n^{\mathrm{BBN}}$ as a function of Weinberg angle $s^2_\mathrm{W}$ for different temperature values with chemical equilibrium neutrino fugacity $\Upsilon_\nu=1$. We can track as a function of $s^2_\mathrm{W}$ the change in value $X^\mathrm{BBN}_n=0.13$ (horizontal black dotted line). The vertical black dotted line marks the adopted CODATA value $s^2_\mathrm{W}=0.223$. The (yellow) shaded domain of $s^2_\mathrm{W}$ is considered in this work.}
\label{WienbergTestfig}
\end{center}
\end{figure}

Before, and at the onset of the BBN period the neutron concentration, without here allowing for nuclear fusion reactions that deplete neutrons, is determined by the ratio of the total WI reaction rate to the Hubble expansion rate, see~\req{RateRatio}. Consequently $X_n$ via correction term depends on magnitude of $G_\mathrm{F}$. It is clear by inspection that the simple freeze-out+lifetime neutron decay model can create an approximation as follows: Taking a high freeze-out condition near $T=2$\,MeV in disagreement with actual values seen in Fig.~\ref{RelaxationRate_fig} we minimize the correction term. The neutron decay shape (compare shape at low $T$) is than a good fit. However, the yield shoulder below $T=30$\,keV will be missed. The results shown in Fig.~\ref{Xn001fig} were obtained without allowing for nuclear fusion reactions which deplete neutron concentration as well.

In the top frame of Fig.~\ref{WienbergTestfig} we show neutron concentration $X_n$ as a function of temperature with different values of Weinberg angle $s^2_\mathrm{W}$. In the bottom frame we show the ratio of the neutron concentration $X_n$ to the standard BBN value $X_n^\mathrm{BBN} = 0.13$ (black dotted line) as a function of the Weinberg angle $s^2_\mathrm{W}$ for different temperatures. At high temperatures, the neutron fraction exhibits only a weak dependence on variations in the weak mixing angle as we expect, since the abundance follows the kinetic detailed balance \lq adiabatic\rq\ result which does not depend on the strength of $G_\mathrm{F}$. In contrast, at lower temperatures the neutron concentration $X_n$ becomes strongly sensitive to the value of the Weinberg angle. The neutron concentration $X_n$ decreases as $s^2_\mathrm{W}$ increases. In particular, near the BBN temperature regime, small variations of the Weinberg angle in the range $0.1<s^2_\mathrm{W}<0.5$ lead to substantial changes in $X_n$, indicating a strong sensitivity. The yellow shaded region marks the range of $s^2_\mathrm{W}$ for which the resummed form $G_\mathrm{F}^\mathrm{sum}$ and the first-order form $G_\mathrm{F}^\mathrm{1st}$ defined below~\req{GFcorr} agree to better than 1\%, identifying the range over which our parametrization of $G_\mathrm{F}(s^2_\mathrm{W})$ is reliable. Outside this region higher-order radiative corrections become important, double counting issue could be relevant.

\begin{figure}[t]
\begin{center}
\hspace*{-0.4cm} \includegraphics[width=5.2in]{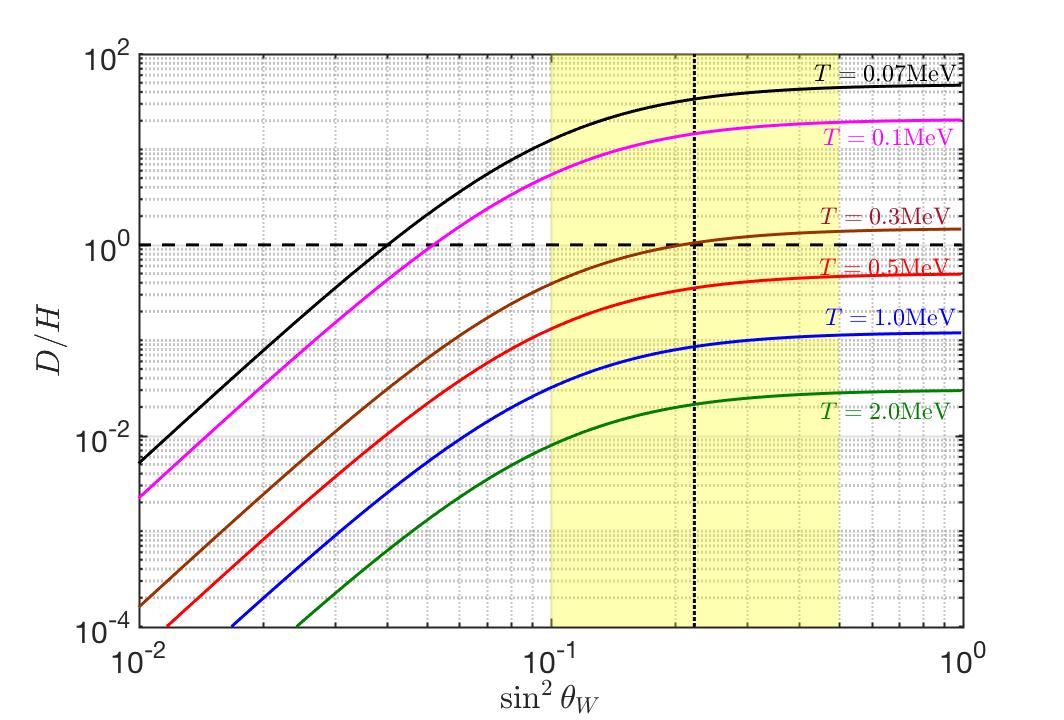}
\caption{The ratio between weak interaction reaction constant $D$,~\req{DimFac} and the Hubble parameter as a function of the Weinberg angle $s^2_\mathrm{W}$ for temperature values $T=2,1,0.5,0.3,0.1,0.07$\,MeV. The horizontal dotted line shows 
$D=H$, and the vertical dotted line marks the Standard Model value $s^2_\mathrm{W}=0.223$. The (yellow) shaded domain of $s^2_\mathrm{W}$ is considered in this work.}
\label{RateRatiofig}
\end{center}
\end{figure}

To better explain the sensitivity of the neutron concentration $X_n$ to the value of $G_\mathrm{F}$, we present in Fig~\ref{RateRatiofig} the ratio of the WI coupling parameter $D$ to the Hubble parameter as a function of Weinberg angle $s^2_\mathrm{W}$ with different temperature values. The results show that near the weak interaction freeze-out temperature, this ratio is sensitive to variations in $s^2_\mathrm{W}$ and thus the neutron concentration at edge of BBN epoch is affected. To accurately capture these effects, it is necessary to study the kinetic evolution of the neutron yield for a range of $s^2_\mathrm{W}$ near to the reference value we adopted $s^2_\mathrm{W}=0.223$. The yellow shaded region marks the range as discussed in prior paragraph.

\section{Conclusions and Discussion}\label{Discussion}
We have described in Section~\ref{Abundance} the competition between the Hubble expansion speed and the kinetic reactions processes responsible for the detailed value of the neutron abundance before and at onset of BBN. The \lq adiabatic limit\rq\ generalizes the thermal approximation. Since the strength of weak interactions, the Fermi constant $G_\mathrm{F}^2$, cancels in the ratio between different electro-weak rates the result is solely dependent on phase space size, this is the detailed balance neutron concentration. The detailed balance differs from thermal equilibrium abundance since we are allowing also for the decay of neutrons. 

To obtain the kinetic neutron concentration $X_n$ we note that several key WI reaction rates seen in Fig.~\ref{RelaxationRate_fig} intersect near to $T$ of interest in BBN. This implies that the neutron freeze-out and decay rates are interwoven with the Hubble expansion rate. The outcome of gradual change of $X_n$ is illustrated for the vacuum SM parameters in Fig.~\ref{Xn001fig}. 

We then proceed to discuss and present in quantitative terms the variation of neutron concentration preceding the temperature regime at edge of BBN as a function of $s^2_\mathrm{W}$, see Fig.~\ref{WienbergTestfig}. It is in this new context that small changes in weak interaction strength driven by Weinberg angle can have impact on neutron abundance available at onset of BBN. We have shown that precision BBN studies that require $X_n$ at the sub-percent level depend on the precise value of $G_\mathrm{F}$, or equivalently---assuming the dominant source of variability is the Weinberg angle---on the value of $s^2_\mathrm{W}$ in the hot primordial plasma at the edge of the BBN epoch.

The results for $X_n$ also depend on precise knowledge of the Hubble expansion parameter. We have been working to develop this understanding for a long time~\cite{Rafelski:2024fej}. We present our sophisticated model as a reference for convenience of the reader in the Appendix~A. In our approach, following on neutrino decoupling, the transfer of entropy and reheating of photons by $e^\pm$-pair annihilation is gradual, obtained from the instantaneous thermal $e^\pm$-pair abundance as a function of temperature. Our precise description has non-negligible improvements.

Looking forward to further work and impact of our results:
\begin{itemize}
\item \textbf{BBN:} In the presence of ongoing BBN nuclear reactions in addition to the neutron loss by decay, neutrons are \lq lost\rq\ to the produced nuclei, for most part in net photo-production of deuterons: Our formulation remains valid in principle, as introducing additional nuclear reaction rate loss (and gain) terms aside of $\Gamma_n$ ($\Gamma_p$) in~\req{RateRatio} we allow for the loss of neutrons due to nuclear reactions such as photoproduction of deuterons, $n+p\to d+\gamma$. The importance of this remark is that the production of neutrons by WI process $\Gamma_{p\to n}$ continues to replenish to some degree such small losses as is implied by the result seen in Fig.~\ref{Xn001fig}. Our detailed WI reaction rates and the underlying numerical data tables are made available so that BBN-network studies can directly incorporate them, providing pre-network neutron-abundance input at the precision level we have established.
\item \textbf{Outlook --- astrophysical neutron-rich environments:} The kinetic-theory framework presented here transfers in principle to other settings where weak rates compete with macroscopic timescales, in particular to core-collapse supernova neutrino spheres, neutron-star mergers, and the early phases of accretion-disk outflows. In all these settings the relevant physics is the competition of in-medium $n\leftrightarrow p$ rates with a hydrodynamic timescale, with the Hubble rate of the present work replaced by the local expansion rate of the flow. The dominant physics in these environments is energy and lepton-number transport rather than precise neutron abundance~\cite{Haxton:2012bk,Janka:2017vlw,Burrows:2020qrp,Cowan:2019pkx}.
\item \textbf{Weinberg symmetry breaking:}
We offered results for $X_n$ motivating future study of unexplored phenomena in electro-weak physics addressing the question if the value of $s^2_\mathrm{W}$ and thus $G_\mathrm{F}$ could be sensitive to environmental factors. In the primordial Universe we suggested the need for the establishment of the temperature dependence of $s^2_\mathrm{W}$ and thus $G_\mathrm{F}$. This clearly requires considerable theoretical effort, far beyond the scope of our work. Without claiming any direct evidence, we have established that within measurement errors of $s^2_\mathrm{W}$, a small residual environmental component could in principle contribute to the discrepancies observed in laboratory neutron-lifetime measurements. Specifically, the observed scatter of measured values of $s^2_\mathrm{W}$ exceeds by a factor 8 the variation required to understand the measured neutron lifetime disagreements.
\end{itemize}

\section*{Author Contributions} This paper was created in collaboration with both authors contributing equally. Both authors have read and agreed to the published version of the manuscript.

\section*{Funding} This research was carried out without sponsor funding.

\section*{Data Availability Statement} The datasets generated during the current study, i.e., data as seen in manuscript figures, are available as ancillary files at arXiv repository, 
\url{https://doi.org/10.48550/arXiv.2603.02652}.

\section*{Acknowledgments} We thank Bernhard Lauss (PSI-Villigen) for valuable comments.

\section*{Conflicts of Interest} The authors declare no conflict of interest.

\vspace{6pt} 

\appendixtitles{no} 
\appendixstart
\appendix
\section[\appendixname~\thesection]{Hubble Parameter and Photon Reheating}\label{Apendix}
In this section, we summarize the key theoretical background relevant to our analysis and in particular the detailed value of the Hubble parameter $H$ in the primordial Universe. Much of the material is based on the review~\cite{Rafelski:2024fej}, which discusses the connection between particle physics and the primordial-Universe cosmology.

In the epoch of interest, $10\,\mathrm{MeV}>T>0.01\,\mathrm{MeV}$, the Universe is dominated by radiation and effectively massless matter behaving like radiation. The Hubble parameter can be written as~
\begin{align}\label{H2g}
H^2=\frac{8\pi G_\mathrm{N}}{3}\left(\rho_\gamma+\rho_{e^\pm}+\rho_{\nu}\right)\,,
\end{align}
where $\rho_i$ is the energy density for $\gamma,e^\pm,\nu$. It is convenient to rewrite the sum of energy density of relativistic species in terms of photo energy density, we have
\begin{align}
&\rho_R=\sum_i\,\rho_i=\frac{\pi^2}{30}g_\ast\,T^4,
\end{align}
where $g_\ast$ counts the effective number of `energy' degrees of freedom, we have
\begin{align}
g_\ast=\!\!\sum_{i=\mathrm{bosons}}g_i&\left({\frac{T_i}{T_\gamma}}\right)^4B+
\frac{7}{8}\!\!\sum_{i=\mathrm{fermions}}g_i\left({\frac{T_i}{T_\gamma}}\right)^4F.
\end{align}
 The functions $B$ and $F$ are defined as
\begin{align}
&B=\frac{15}{\pi^4}\int^\infty_{m_i/T}\,dx\sqrt{x^2-\left(\frac{m_i}{T}\right)^2}\frac{x^2}{e^x-1},\\
&F=\frac{8}{7}\times\frac{15}{\pi^4}\int^\infty_{m_i/T}\,dx\sqrt{x^2-\left(\frac{m_i}{T}\right)^2}\frac{x^2}{e^x+1}.
\end{align}
\begin{figure}[t]
\begin{center}
\hspace*{-0.4cm} \includegraphics[width=5.2in]{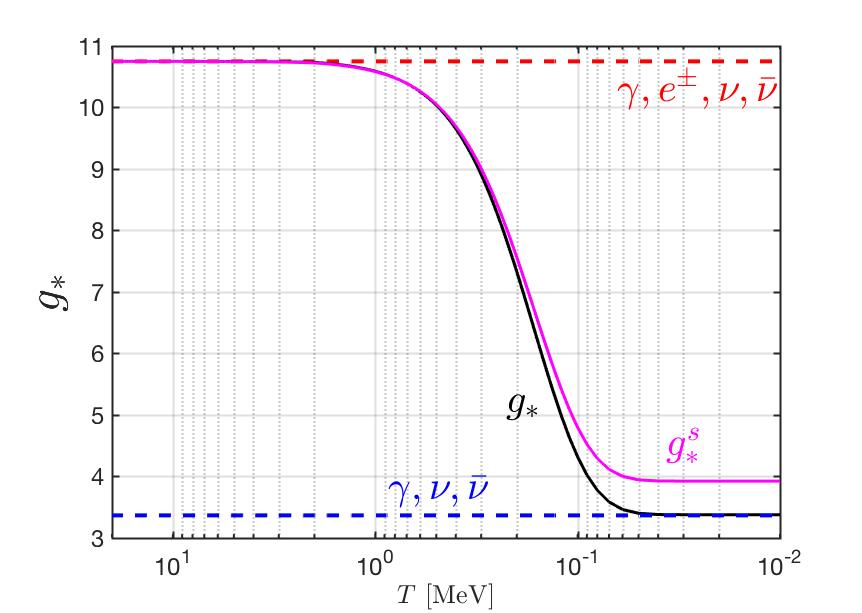}
\caption{The effective number of `energy' degrees of freedom $g_\ast$ as a function of temperature. The red dashed line represents the degrees of freedom from photon, neutrino, and massless electron/positron. The blue dashed line labels the degree of freedom for photon, and cold neutrino.}
\label{EnergyDOF_fig}
\end{center}
\end{figure}
In Fig.~\ref{EnergyDOF_fig}, we show the effective number of energy degrees of freedom $g_\ast$ as a function of temperature. The red dashed line shows the contribution from photons, neutrinos, and massless electrons/positrons, while the blue dashed line represents the contribution from photons and cold neutrinos. 
\begin{figure}[t]
\begin{center}
\hspace*{-0.4cm} \includegraphics[width=5.2in]{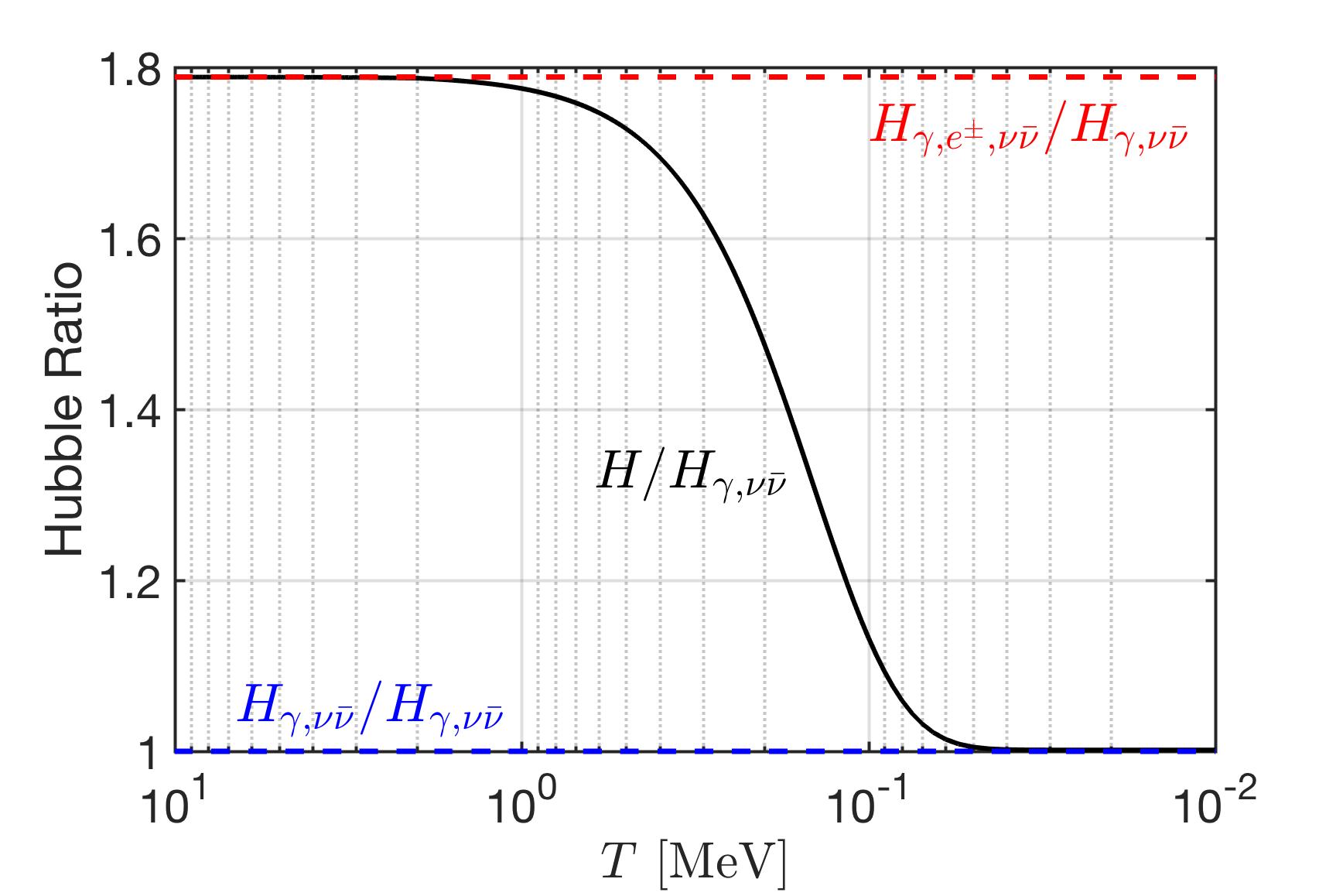}
\caption{The ratio between Hubble parameter as a function of temperature from $10\geqslant T\geqslant0.01$MeV in the primordial Universe. It shows that the transition between Hubble parameter from contributions by photons, neutrinos, and massless electrons/positrons to contributions from only photons and neutrinos is driven by the smooth disappearance of nonrelativistic $e^\pm$.}
\label{Hubble_fig}
\end{center}
\end{figure}
In Fig.~\ref{Hubble_fig}, we plot the Hubble parameter as a function of temperature from $10\geqslant T\geqslant0.01$MeV in the primordial Universe. It shows that the transition between Hubble parameter from contributions by photons, neutrinos, and massless electrons/positrons to contributions from only photons and neutrinos is driven by the smooth disappearance of nonrelativistic $e^\pm$.


After neutrino freeze-out and when $m_e\gg T$, the $e^{\pm}$ becomes non-relativistic and annihilate. In this case, their entropy is transferred to the other relativistic particles still present in the cosmic plasma, i.e., photons, resulting in an increase in photon temperature as compared to the freestreaming neutrinos. From entropy conservation we have
\begin{align}
\label{Entropy}
\frac{2\pi}{45}g^s_\ast(T_k)T^3_kV_k+S_{\nu}(T_k)=\frac{2\pi}{45}g^s_\ast(T)T^3V+S_{\nu}(T),
\end{align}
where we use the subscripts $k$ to denote quantities for neutrino freeze-out and $g^s_\ast$ counts the degree of freedom for relativistic species in primordial Universe. The effective number of `entropy' degrees of freedom.can be written as
\begin{align}
g^s_\ast\!=\!\!\sum_{i=\mathrm{bosons}}\!\!g_i&\left({\frac{T_i}{T_\gamma}}\right)^3\!B^s+\frac{7}{8}\!\!\sum_{i=\mathrm{fermions}}\!\!g_i\left({\frac{T_i}{T_\gamma}}\right)^3\!F^s,
\end{align}
where the functions $B^s(m_i/T)$ and $F^s(m_i/T)$ are defined as
\begin{align}
&B^s\!=\!\frac{45}{12\pi^4}\!\!\int^\infty_{m_i/T}\!\!\!\!\!\!dx\frac{\sqrt{x^2\!-\!\left({m_i}/{T}\right)^2}\left[4x^2\!-\!\left({m_i}/{T}\right)^2\right]}{e^x-1},\\
&F^s=\frac{8}{7}\!\!\times\!\!\frac{45}{12\pi^4}\int^\infty_{m_i/T}\!\!\!\!\!dx\frac{\sqrt{x^2\!-\!\left(\frac{m_i}{T}\right)^2}\left[4x^2\!-\!\left(\frac{m_i}{T}\right)^2\right]}{ e^x+1}.
\end{align}

After neutrino freeze-out, their entropy is conserved independently and the temperature can be written as
\begin{align}
T_\nu\equiv\frac{a(t_k)}{a(t)}T_k=\left(\frac{V_k}{V}\right)^{1/3}T_k.
\end{align}
In this case, from entropy conservation, Eq.\,(\ref{Entropy}), we obtain the neutrino temperature
\begin{align}
\label{Neutrino_Photon}
T_\nu=\frac{T}{\kappa},\,\,\,\,\,\,\kappa\equiv\left[\frac{g^s_\ast(T_k)}{g^s_\ast(T)}\right]^{1/3}.
\end{align}

For neutrinos, after neutrino/antineutrino kinetic freeze-out they become free streaming particles. If we assume that kinetic freeze-out occurs at some time $t_k$ and temperature $T_k$, then for $t>t_k$ the free streaming distribution function can be written as~\cite{Birrell:2012gg}
\begin{align}
f_{\nu}=\frac{1}{\Upsilon_\nu^{-1}\exp{\left(\sqrt{\frac{E^2-m_\nu^2}{T_\nu^2}+\frac{m^2_\nu}{T^2_k}}+\frac{\mu_{\bar{\nu}}}{T_k}\right)+1}},
\end{align}
for antineutrinos. In our calculation, we assume the condition $T_k=2\,\mathrm{MeV}\gg\mu_{\bar{\nu}},\,m_\nu$, i.e., we consider the massless neutrino in plasma.

The relation between temperature and time follows from the isentropic expansion of the photon-coupled plasma, $g^s_\ast(T)\,T^3a^3=\mathrm{const}$ after neutrino decoupling, see Section~1.4.2 and Eqs.~(1.58)--(1.59) in Ref.\,\cite{Rafelski:2024fej}:
\begin{align}\label{dtdT}
\frac{dt}{dT}=-\frac{1}{H\,T}\left(1+\frac{1}{3}\frac{d\ln g^s_\ast}{d\ln T}\right).
\end{align}
The derivative term accounts for the slower cooling of photons while $e^\pm$ annihilate. The time axes of Figs.~\ref{RelaxationRate_fig} and~\ref{Xn001fig} and the kernel~\req{RateRatio} use this relation.


\reftitle{References}






\end{document}